\documentclass[%
 reprint,
 amsmath,amssymb,
 aps,
]{revtex4-2}

\usepackage{graphicx}% Include figure files
\usepackage{dcolumn}% Align table columns on decimal point
\usepackage{bm}% bold math
\usepackage{empheq}
\usepackage{color}
\usepackage{wrapfig}
\usepackage{hyperref}% add hypertext capabilities
\newtheorem{remark}{Remark}

\begin{document}

\preprint{APS/123-QED}

\title{Boundary accumulations of active rods in  microchannels \\with elliptical cross-section}% Force line breaks with \\
%\thanks{A footnote to the article title}%

\author{Chase Brown}
% \altaffiliation[Also at ]{Department of Mathematics \\ University of California, Riverside\\
% 900 University Ave.\\
%Riverside, CA 92521}%Lines break automatically or can be forced with \\

\affiliation{%
 Department of Mathematics \\ University of California, Riverside\\
 900 University Ave.\\
Riverside, CA 92521
}%

%\collaboration{MUSO Collaboration}%\noaffiliation

\author{Shawn D. Ryan}
 \homepage{http://academic.csuohio.edu/ryan-shawn/}
\affiliation{
 Department of Mathematics and Statistics
}%
\affiliation{
 Center for Applied Data Analysis and Modeling\\
 Cleveland State University\\
 Cleveland, OH 44115
}%

\author{Mykhailo Potomkin}
 \homepage{Corresponding Author}
\affiliation{%
 Department of Mathematics \\ University of California, Riverside\\
 900 University Ave.\\
Riverside, CA 92521
}%

%\collaboration{CLEO Collaboration}%\noaffiliation

\date{\today}% It is always \today, today,
             %  but any date may be explicitly specified

\begin{abstract}
Many motile microorganisms and bio-mimetic micro-particles have been successfully modeled as active rods - elongated bodies capable of self-propulsion. A hallmark of active rod dynamics under confinement is their tendency to accumulate at the walls. Unlike passive particles, which typically sediment and cease their motion at the wall, accumulated active rods continue to move along the wall, reorient, and may even escape from it. The dynamics of active rods at the wall and those away from it result in complex and non-trivial distributions. In this work, we examine the effects of wall curvature on active rod distribution by studying elliptical perturbations of tube-like microchannels, that is, the cylindrical confinement with a circular cross-section, common in both nature and various applications. By developing a computational model for individual active rods and conducting Monte Carlo simulations, we discovered that active rods tend to concentrate at locations with the highest wall curvature. We then investigated how the distribution of active rod accumulation depends on the background flow and orientation diffusion. Finally, we used a simplified mathematical model to explain why active rods preferentially accumulate at high-curvature locations.

%\begin{description}
%\item[Usage]
%Secondary publications and information retrieval purposes.
%\item[Structure]
%You may use the \texttt{description} environment to structure your abstract;
%use the optional argument of the \verb+\item+ command to give the category of each item. 
%\end{description}
\end{abstract}

%\keywords{Suggested keywords}%Use showkeys class option if keyword
                              %display desired
\maketitle

%\tableofcontents

\section{\label{sec:level1}Introduction}
%%%%%%%%%%%%%%%%%%%%%%%%%%%%%%%%%%%%%%%%%%%%%%%%%%%%%%%%%%
% 1. Many living systems are active matter 
% Self-propelled rod models
%%%%%%%%%%%%%%%%%%%%%%%%%%%%%%%%%%%%%%%%%%%%%%%%%%%%%%%%%%
Systems of motile agents capable of persistent directed motion fall under the category of active matter. At micro-scale, such systems can be exemplified by both biological systems, like suspensions of motile bacteria \cite{aranson2022bacterial}, and biomimetic ones, such as those containing chemically driven bimetallic particles \cite{PaxSenMal05,PaxSunMalSen06,PaxBakKliWanMalSen06}. Active matter displays many fascinating phenomena related to collective behavior such as large-scale fluid dynamics and flocking \cite{wensink2012meso,SokAra2012,SokGolFelAra2009,gompper20202020}. The challenge in studying such systems is that they are in fact non-equilibrium by their continuous injection of energy into the system through self-propelled motion of autonomous components \cite{peruani2020review,hallatschek2023proliferating}.  This injection of energy can lead to counter-intuitive results such as decreased viscosity \cite{sokolov2009reduction,HaiSokAraBer09,RyaHaiBerZieAra11,potomkin2016ev}, movement against the background flow \cite{HilKalMcMKos2007,fu2012bacterial,PalSacAbrBarHanGroPinCha2015,YuaRaiBau2015,ren2017rheotaxis,rheotaxis2018baker}, and the formation of whirls and jets \cite{WuLib2000,VisZaf2012,RyaSokBerAra13}.  Also, fascinating is the multiscale natural and universal properties of active matter ranging from nanoscale to human-scale \cite{gompper20202020}.  All these emergent non-Newtonian macroscopic behaviors have intrigued researchers for some time and inspired the development of novel mathematical models to study the underlying behavior.  For a complete overview of the field of active matter we recommend consulting a few recent review articles \cite{hallatschek2023proliferating,gompper20202020,aranson2022bacterial,vrugt2024review}. 

%%%%%%%%%%%%%%%%%%%%%%%%%%%%%%%%%%%%%%%%%%%%%%%%%%%%%%%%%%
% 2. Active matter in confined environments - wall accumulation and rheotaxis 
%%%%%%%%%%%%%%%%%%%%%%%%%%%%%%%%%%%%%%%%%%%%%%%%%%%%%%%%%%
While the field of active matter is vast and far-reaching as well as covering many scales; herein, we narrow our focus to the study of self-propelled ``active" rods in confined environments.  An active rod, a finite directed segment experiencing a constant propulsion force in the direction its facing, has proven to be an efficient model for elongated micro-swimmers that are constituents of an active matter system \cite{potomkin2017focusing,rubio2021self,peruani2020review}. Two distinguishing features of active swimming at confinement, successfully captured by the active rod model, are {\it bordertaxis}, which is the tendency of active swimmers to accumulate at boundaries, and {\it negative rheotaxis}, which refers to upstream swimming. 

The wall accumulation can result from different factors such as hydrodynamic attraction by the wall \cite{drescher2011fluid} or electrostatic interactions \cite{ren2017rheotaxis}. However, the primary reason for the rod's accumulation is its activity. In other words, an active rod confined between boundaries, in the absence of mechanisms to avoid them (such as wall repulsion), will inevitably collide with the boundaries over time. Bordertaxis does not require a background flow, and the accumulation of active rods can be observed in an ambient environment. In contrast, negative rheotaxis describes how an active rod responds to a background flow. To explain negative rheotaxis, note that as an active rod approaches a wall, the background flow reorients it upstream. This occurs because of the no-slip boundary conditions: the background flow vanishes at the wall and increases as one moves away from it. Consequently, the drag force exerted by the background flow on the front of the active rod is lower than on the rear, resulting in a torque that reorients the active rod to face the current, thereby exhibiting negative rheotaxis. Combination of these two phenomena have been mathematically analyzed for simple boundary geometries, for example, in \cite{potomkin2017focusing,rubio2021self} for nozzles, and \cite{EzhSai2015} for a chamber.   

There is a rich array of dynamics displayed by active rods near walls that is still not well understood. After colliding with a boundary, accumulated active rods do not simply settle but instead follow the boundary and may even escape from it. Thus, active rod dynamics is a non-trivial combination of swimming in the fluid bulk  and on the boundary surface. The geometry of the boundary wall can significantly affect the active rod distribution and properties of their individual trajectories. The most ubiquitous geometries in applications are cylindrical microchannels with circular or rectangular cross-sectional geometry. It is known, for example, that bacteria in microchannels with rectangular cross-sections tend to accumulate at corners forming ``four-lane" transport \cite{HilKalMcMKos2007,nuris2015,Bukatin2020}. It is natural to hypothesize that corners and locations of high or low curvature may serve to control the transport of micro-swimmers such as bacteria, sperms, and bio-mimetic particles. For example, such a control would be relevant in addressing issues of bacterial contamination due  to their upstream swimming, the reason of
catheter-associated urinary tract infection, one of the most common type acquired in health care
facilities \cite{nicolle2014}, and formation of hazardous bacteria films in channel geometries \cite{park2004}. Our work is further motivated by a recent experimental finding how bacteria can utilize the upstream swimming near corners and crevices to invade and contaminate microchannels, the effect called ``super-contamination" \cite{figueroa2020coli}. The term comes from the idea that bacteria could migrate upstream back to a source and thus contaminating it.   
Though it seems counter-intuitive that bacteria have the ability to have a net progress upstream given the run-and-tumble dynamics of bacteria, mathematical modeling can be used to reveal the underlying dynamics behind these observations.         

In this work, we seek to study the effect of cross-sectional geometry on the distributions of active rods inside the three-dimensional microchannel and statistical properties of their individual trajectories. Our focus is on how the distribution of an individual active rod depends on the boundary curvature. To this end, we consider an elliptical cross-section, so that the curvature varies along the cross-section.   
Furthermore, we illuminate the effect of key parameters such as the flow rate, rotational diffusion, and aspect ratio of the cross-section in the channels on the macroscopic behavior of active rod distribution and the effect that varying curvature has on the distribution. We use the mathematical model developed herein to investigate through histograms of active rod distributions in the cross-section. This investigation lays the groundwork for further theoretical and experimental approaches toward complete understanding of geometric control of bacterial contamination.
%%%%%%%%%%%%%%%%%%%%%%%%%%%%%%
%%%%% What to mention in intro

% - why elliptic? To study the effect of curvatures, Make a huge difference compared to flat surfaces & perturbation of circular shapes  
% - Prototypical model of active rods, Real propulsion is complicated + Hydrodynamics Interactions + Propulsion at wall may be different from propulsion in bulk

\section{Method}
We generalize the model from \cite{potomkin2017focusing} to study the dynamics of active rod-shaped swimmers moving in a cylindrical microchannel with various cross-sections. Three main components of the model described below are the dynamics of an active rod in the fluid, the fluid flow vector field, and the active rod's dynamics at the microchannel wall.     
%%%%%%%%%%%%%%%%%%%%%%%%%%%%%%%%%%%%%%%%%%
% Here, we can add that 
% - we study the statistical properties of individual active rod trajectories. No interactions, mention Fokker-Planck
% - No interaction with walls. No collision paradox

%%%%%%%%%%%%%%%%%%%%%%%%%%%%%%%%%%%%%%%%%%%%%%%%%%%%%%%%%%
% 1. Model of a rod in bulk
%%%%%%%%%%%%%%%%%%%%%%%%%%%%%%%%%%%%%%%%%%%%%%%%%%%%%%%%%%
\begin{figure}[t]
\includegraphics[width=0.25\textwidth]{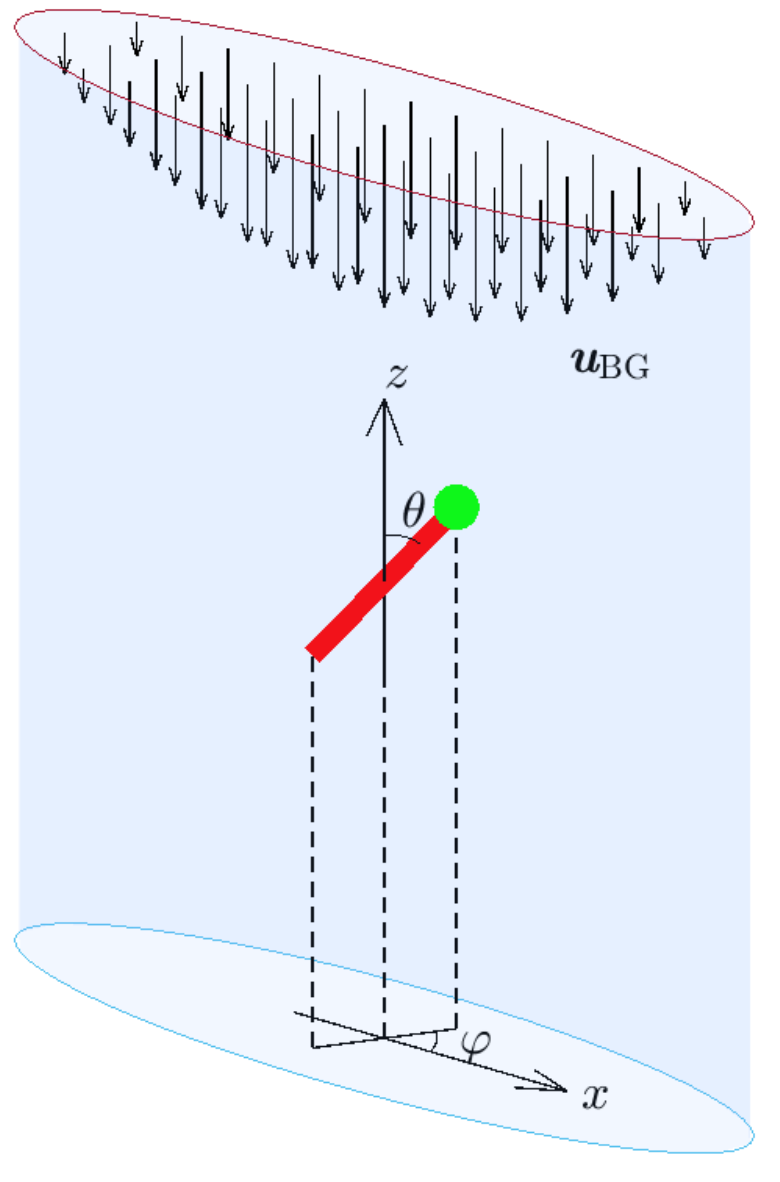}
\caption{{\bf Illustration for the active rod model.} The rod is depicted as a red segment with the green front pointing towards the propulsion direction. Black arrows on the top show fluid flow field $\boldsymbol{u}_{\text{BG}}$. }
\label{fig:1}
\end{figure}

The state of an active rod is uniquely determined by its center location $\boldsymbol{r}(t)=(x(t),y(t),z(t))^{\text{T}}$ and the unit orientation vector \begin{equation}
\boldsymbol{p}(t)=\left[\begin{array}{c}
\cos(\varphi(t))\sin(\theta(t))\\\sin(\varphi(t))\sin(\theta(t))\\\cos(\theta(t))\end{array}\right].
\label{eqn:def_of_p}
\end{equation}
The orientation vector $\boldsymbol{p}$ defines the rod's front direction with orientation angles $0\leq \theta \leq \pi$ and $-\pi\leq \varphi < \pi$, see Figure~\ref{fig:1}. 

The dynamics of the active rod is calculated from force and torque balances. Namely, assume that the fluid exerts the force $-\eta \boldsymbol{V}\,\text{d}s$ onto element $\text{d}s$ of the rod whose relative velocity with respect to the background flow is \begin{equation}
\boldsymbol{V}(s)=\frac{\text{d}}{\text{d}t}\left[\boldsymbol{r}+s\boldsymbol{p}\right]-\boldsymbol{u}|_{\boldsymbol{r}+s\boldsymbol{p}}.
\end{equation}
Here, $s$ is the natural parameter of the rod of length $\ell$, so $-\ell/2\leq s \leq \ell/2$ and vector-function $\boldsymbol{u}(\boldsymbol{r})$ describes the background flow and solves Stokes equation.   Using the smallness of the rod's length compared to the width of the microchannel, we can approximate the expression for the relative velocity by 
\begin{equation}
\boldsymbol{V}(s)\approx \dot{\boldsymbol{r}}-\boldsymbol{u}+s\left(\dot{\boldsymbol{p}}-(\nabla\boldsymbol{u})\boldsymbol{p}\right),\label{eqn:approx}
\end{equation}
where $\boldsymbol{u}$ and $\nabla \boldsymbol{u}$ are computed at $\boldsymbol{r}$. Notations $\dot{\boldsymbol{r}}$ and $\dot{\boldsymbol{p}}$ stand for time-derivatives of ${\boldsymbol{r}}$ and ${\boldsymbol{p}}$, respectively.
 Parameter $\eta$ is related to the viscosity of the background fluid. 

The force and torque balances for the active rod are written as
\begin{eqnarray}
\boldsymbol{0} &=& -\eta \int\limits_{-\ell/2}^{\ell/2}\boldsymbol{V}(s)\,\text{d}s + F_{\text{prop}}\boldsymbol{p},\label{eqn:fb}\\
\boldsymbol{0} &=& -\eta \int\limits_{-\ell/2}^{\ell/2}s\boldsymbol{p}\times\boldsymbol{V}(s)\,\text{d}s + \boldsymbol{p}\times\sqrt{2D_{\text{rot}}}\dot{\boldsymbol{W}}.\label{eqn:tb}
\end{eqnarray}
The force balance \eqref{eqn:fb} means that the rod moves because of the background flow and the propulsion of the given strength $F_{\text{prop}}$ in the direction of the rod's orientation $\boldsymbol{p}$. The torque balance \eqref{eqn:tb} describes how the rod is reoriented due to the background flow and noise through the random orientation governed by the rotational diffusion $D_{\text{rot}}$. The random force reorienting the rod without displacing it is of the form $\dot{\boldsymbol{W}} = \dot{W}_{\varphi}\boldsymbol{e}_\varphi+\dot{W}_{\theta}\boldsymbol{e}_{\theta}$, where 
\begin{equation*}
\boldsymbol{e}_\varphi = \left[
\begin{array}{c}
-\sin(\varphi(t))\\
\cos(\varphi(t))\\
0
\end{array}\right], \,
\boldsymbol{e}_\theta = \left[
\begin{array}{c}
\cos(\varphi(t))\cos(\theta(t))\\
\sin(\varphi(t))\cos(\theta(t))\\
-\sin(\theta(t))
\end{array}\right].
\end{equation*}
Note that the triple $(\boldsymbol{p},\boldsymbol{e}_{\varphi},\boldsymbol{e}_{\theta})$ is an orthonormal basis. 

Next, we rewrite equations \eqref{eqn:fb} and \eqref{eqn:tb}  as a system for $\boldsymbol{r}$ and $\boldsymbol{p}$. To this end, use the expression \eqref{eqn:approx}, take the cross product of both sides of \eqref{eqn:tb} with the vector $\boldsymbol{p}$, and apply the BAC-CAB rule implying 
$\boldsymbol{p}\times(\boldsymbol{p}\times \boldsymbol{\tau})=-\boldsymbol{\tau}$ for $\boldsymbol{\tau}\cdot \boldsymbol{p}=0$, and $\boldsymbol{p}\times(\boldsymbol{p}\times \boldsymbol{v})=-(\text{I}-\boldsymbol{p}\boldsymbol{p}^{\text{T}})\boldsymbol{v}$ for an arbitrary vector $\boldsymbol{v}$. We obtain
\begin{subequations} \label{eqn:system}
\begin{empheq}[left=\empheqlbrace]{align}
&\dot{\boldsymbol{r}}=\boldsymbol{u}+v_{\text{prop}}\boldsymbol{p}\label{subeqn:r}\\
&\dot{\boldsymbol{p}}=\left(\text{I}-\boldsymbol{p}\boldsymbol{p}^{\text{T}}\right)(\nabla \boldsymbol{u})\boldsymbol{p}+\sqrt{2\hat{D}_{\text{rot}}}\dot{\boldsymbol{W}}\label{subeqn:p}
\end{empheq}
\end{subequations}
where $v_\text{prop}=(\eta \ell)^{-1}F_\text{prop}$ is the propulsion speed and $\hat{D}_{\text{rot}}={144 D_{\text{rot}}}/{\ell^6\eta^2}$ is the the effective rotation diffusion coefficient.

It is convenient to rewrite the vector equation \eqref{subeqn:p} as a system for orientation angles $\theta$ and $\varphi$. Use the following identity that can be obtained by differentiating \eqref{eqn:def_of_p}: 
\begin{equation}
\dot{\boldsymbol{p}}=\dot{\varphi}\sin(\theta) \boldsymbol{e}_\varphi + \dot\theta\boldsymbol{e}_{\theta}.
\end{equation}
By using this identity and taking dot products of both sides of equation \eqref{subeqn:p} with $\boldsymbol{e}_{\varphi}$ and $\boldsymbol{e}_{\theta}$ we obtain equations for $\varphi$ and $\theta$, respectively: 
\begin{equation}
\left\{
\begin{array}{l}
\dot{\theta} = (\nabla \boldsymbol{u})\boldsymbol{p}\cdot \boldsymbol{e}_{\theta} + \sqrt{2\hat{D}_{\text{rot}}}\dot{W}_\theta\\
\dot{\varphi} = (\sin(\theta))^{-1}\left[(\nabla \boldsymbol{u})\boldsymbol{p}\cdot \boldsymbol{e}_{\varphi}+ \sqrt{2\hat{D}_{\text{rot}}}\dot{W}_\varphi\right]
\end{array}
\right. \label{eqn:angles}
\end{equation}
Equations \eqref{subeqn:r} and \eqref{eqn:angles} describe the active rod's dynamics in the background fluid flow $\boldsymbol{u}$. 

%%%%%%%%%%%%%%%%%%%%%%%%%%%%%%%%%%%%%%%%%%%%%%%%%%%%%%%%%%
% 2. Background flow - Reduction to Poisson equation
%%%%%%%%%%%%%%%%%%%%%%%%%%%%%%%%%%%%%%%%%%%%%%%%%%%%%%%%%%
To compute the background fluid flow $\boldsymbol{u}$, one needs to solve Stokes equation: 
\begin{equation}
\left\{
\begin{array}{l}
-\mu \Delta \boldsymbol{u} + \nabla \Pi  = 0,\\
\hspace{15pt}\nabla\cdot \boldsymbol{u} = 0, \\
\hspace{15pt}\boldsymbol{u}|_{\partial M} = 0.
\end{array}
\right.\label{eqn:stokes}
\end{equation}
Here, $\Pi$ is the pressure and the first two equations hold inside the microchannel $M$ with the boundary wall $\partial M$. The third equation in \eqref{eqn:stokes} is the no-slip boundary condition at the wall $\partial M$. Cylindrical geometry of the microchannel allows for simplification of \eqref{eqn:stokes}. Namely, let $z$-axis be the major axis of the cylinder, that is, $M=\{(x,y,z):(x,y)\in\Omega,-\infty<z<\infty\}$ where $\Omega$ is the two-dimensional domain indicating the cross-section geometry. Then \eqref{eqn:stokes} admits a solution in the form $\boldsymbol{u}=w(x,y)\boldsymbol{e}_z$ and $\Pi=[\Pi]z$ where $\boldsymbol{e}_z=(0,0,1)$ and $[\Pi]$ is a constant quantifying the pressure gradient along the microchannel. This solution type is relevant for our purposes since it describes a pressure-driven flow along the microchannel. The equation for $w(x,y)$ reduces to a two-dimensional Poisson's equation with a constant in the right-hand side. Specifically, $w(x,y)$ solves the boundary value problem:  
\begin{equation}
\left\{
\begin{array}{l}
\Delta_{x,y} w =[\Pi]/\mu \,\text{ in }\Omega,\\
w = 0\,\text{ on }\partial \Omega.
\end{array}
\right.
\label{eqn:bvp}
\end{equation}
This equation has an analytical solution for some geometries $\Omega$ such as an ellipse or a rectangle. Background flows $\boldsymbol{u}$ based on these analytical solutions $\hat{w}$ are known as Poiseuille flows.
For example, if $\Omega$ is the ellipse $x^2/a^2+y^2/b^2=1$, then 
\begin{equation*}
w(x,y) = \dfrac{a^2b^2[\Pi]}{2\mu(a^2+b^2)}\left[\dfrac{x^2}{a^2}+\dfrac{y^2}{b^2} - 1\right].
\end{equation*}
%If $\Omega$ is the rectangular $[-a,a]\times[-b,b]$, then 
%\begin{equation*}
%\resizebox{0.5\textwidth}{!}{$\hat{w}(x,y) = -\dfrac{16b^2}{\pi^3}\sum\limits_{n=1}^{\infty}\dfrac{(-1)^{n+1}}{(2n-1)^3}\left\{1-\dfrac{\cosh\left(\dfrac{(2n-1)\pi}{2b}x\right)}{\cosh\left(\dfrac{(2n-1)\pi}{2b}a\right)}\right\}\cos\left(\dfrac{(2n-1)\pi y}{2b}\right).$}
%\end{equation*}
The constant $[\Pi]/\mu$ is related to the volume of the fluid passing through a given cross-section per unit of time $\nu_e$, that is, 
\begin{equation*}
\nu_e=\int\limits_{\Omega} |w(x,y)|\,\text{d}x\text{d}y = \dfrac{[\Pi]\kappa}{\mu}, 
\end{equation*}
where the parameter $\kappa = \int\limits_{\Omega} |\hat{w}(x,y)|\,\text{d}x\text{d}y$ depends on geometry of $\Omega$ only. Here, $\hat{w}$ is the solution of the boundary-value problem \eqref{eqn:bvp} with $1$ in place of $[\Pi]/\mu$. For the elliptical cross-section, we have that $\kappa = \int\limits_{\Omega}|\hat{w}|\,\text{d}x\text{d}y=\pi a^3b^3/(4(a^2+b^2))$. As it was pointed in \cite{potomkin2017focusing}, it is convenient to express the flow rate in terms of the flow speed at the microchannel's center-line $x=y=0$, which is related to $\nu_e$ through the following equality: 
\begin{equation}
\nu_e = \dfrac{\pi a b}{2}|w(0,0)|.
\label{eq:express-flow}
\end{equation}

We formulate boundary conditions for active rods interacting with confinement. We impose that an active rod cannot penetrate the boundary but still can move tangentially along the boundary and reorient. That is, the equation for $\boldsymbol{r}$ becomes 
\begin{equation}
\dot{\boldsymbol{r}} = (\text{I}-q\boldsymbol{n}\boldsymbol{n}^{\text{T}})(\boldsymbol{u}+v_{\text{prop}}\boldsymbol{p}+\sqrt{2\hat{D}_{\text{rot}}}\dot{\boldsymbol{W}}), \label{eqn:constraint} 
\end{equation}
where $q=0$ if the rod is in the bulk and $q=1$ if the rod is at the confinement; vector $\boldsymbol{n}$ is the unit outward normal vector for $\Omega$. The numerical implementation of this boundary condition is subtle since the rod is never exactly on the wall. 
It is often the case that when a numerical step is applied to the system \eqref{eqn:system}, the rod's front $\boldsymbol{r}_f = \boldsymbol{r}+\ell/2 \boldsymbol{p}$ ends up outside the wall. If it happens, we adjust the rod location $\boldsymbol{r}$ by shifting it along $-d\boldsymbol{n}$ where $d$ is the distance between the $\boldsymbol{r}_{f}$ and the wall and $\boldsymbol{n}$ is the unit outward normal vector computed at the projection of $\boldsymbol{r}_f$ onto the wall, $\boldsymbol{r}_{W}=(a\cos(\alpha_W),b\sin(\alpha_W),z_f)$. Here, $z_f$ is the $z$-coordinate of $\boldsymbol{r}_f=(x_f,y_f,z_f)$, and the parameter $\alpha_W$ is the solution of the minimization condition of $\min\limits_{0\leq\alpha_W<2\pi}\|\boldsymbol{r}_f-\boldsymbol{r}_W(\alpha_{W})\|^2$:
\begin{equation*}
\resizebox{0.475\textwidth}{!}{$(a^2-b^2)\cos(\alpha_W)\sin(\alpha_W)-ax_f\sin(\alpha_W)+by_f\cos(\alpha_W)=0.$}
\end{equation*}
After this adjustment, the rod is located inside the computational domain $\Omega$ with its front on the wall. 

We note that  propulsion mechanisms of specific types of micro-swimmers ({\it e.g.,} sperm, motile bacteria {\it E. coli} or {\it B. subtilis}, and bimetallic rods) are highly diverse and complex. In studying macroscopic properties of active swimmers, such as dynamics of their distributions, it is common to represent propulsion mechanism as a constant force in the direction of $\boldsymbol{p}$ and not bother with the details of self-propulsion generation. However, confinement, besides being an obstacle, are known to alter the micro-swimmer dynamics \cite{RamTulPha1993,diluzio2005}, for example, by hydrodynamic attraction or affecting beating of flagella, a primary means of propulsion for many motile bacteria. Here, our focus is on prototypical active rods that maintain self-propulsion constant away and at the wall, and the wall is merely a geometric constraint expressed by the active rod boundary condition \eqref{eqn:constraint}. We note that numerical results in \cite{rubio2021self} based on these modeling assumptions showed good agreement with experiments on bimetallic rods. Generally, any discrepancies between experimental observations for specific micro-swimmers and results of the active rod model described above may highlight particular features of these micro-swimmers.  
Nevertheless, we would like to study the effect of the wall torque which was not considered in earlier works \cite{potomkin2017focusing,rubio2021self}. Namely, according to Newton's third law, the active rod experiences a reaction force that balances its self-propulsion, preventing the active rod penetrating the wall. This force has been taken into account by the term with parameter $q$ in the force balance \eqref{eqn:constraint}. But this force is applied to the front edge of the active rod and thus introduces a torque that must affect the equation for $\boldsymbol{p}$. Namely, to take into account this torque we need to add the following term to the right-hand side of the equation \eqref{subeqn:p}:
\begin{equation}
-q\mathcal{T}\dfrac{6v_{\text{prop}}}{\ell}(\boldsymbol{n}^{\text{T}}\boldsymbol{p})(\text{I}-\boldsymbol{p}\boldsymbol{p}^{\text{T}})\boldsymbol{n}.
\label{eqn:wall-torque}
\end{equation}
Here, $\boldsymbol{n}$ is the outward normal vector where the active rod's front edge $\boldsymbol{r}_f$ touches the wall. The parameter $\mathcal{T}$ is such that if $\mathcal{T}=0$, then the wall torque is neglected and if $\mathcal{T}=1$, then the wall torque is applied. We also can vary $0\leq \mathcal{T}\leq 1$ to study the role of the wall torque. The derivation of \eqref{eqn:wall-torque} is relegated to Supplementary Information. It was shown that the wall torque may help bacteria to escape  from wall entrapment  \cite{potomkin2017flagella}. In that study, bacteria are initially passive and attached to the wall. However, after nutrients are added to the suspension, the bacteria become active and start rotating at the wall due to the wall torque, eventually escaping from it.

\medskip 

\section{Results}

%%%%%%%%%%%%%%%%%%%%%%%%%%%%%%%%%%%%%%%%%%%%%%%%%%%%%%%%%%
% 1. Ellipse
% (a) Concentration at high curvature 
% I. Video for flow rate 1: horizontal 3D channel in the first row
% xy-cross-section and plot for fraction of accumulated rods in the second row
% II. Video Comparison of three flow rates. 
% III. Angular distribution

% TODO LIST
% 1. MAKE SURE THAT IT IS CLEAR THAT RODS DO NOT INTERACT. THIS IS ABOUT MONTE-CARLO SIMULATIONS
% 2. MAKE SURE THAT ALL PARAMETERS ARE INDICATED FOR SIMULATION RESULTS IN THIS SECTION
% 3. (MAYBE IN DISCUSSION) WHAT ARE VALUES OF PARAMETERS IN REAL LIFE (BACTERIA, SPERM, AU-PT PARTICLES)

\begin{figure}[t]
\includegraphics[width=0.5\textwidth]{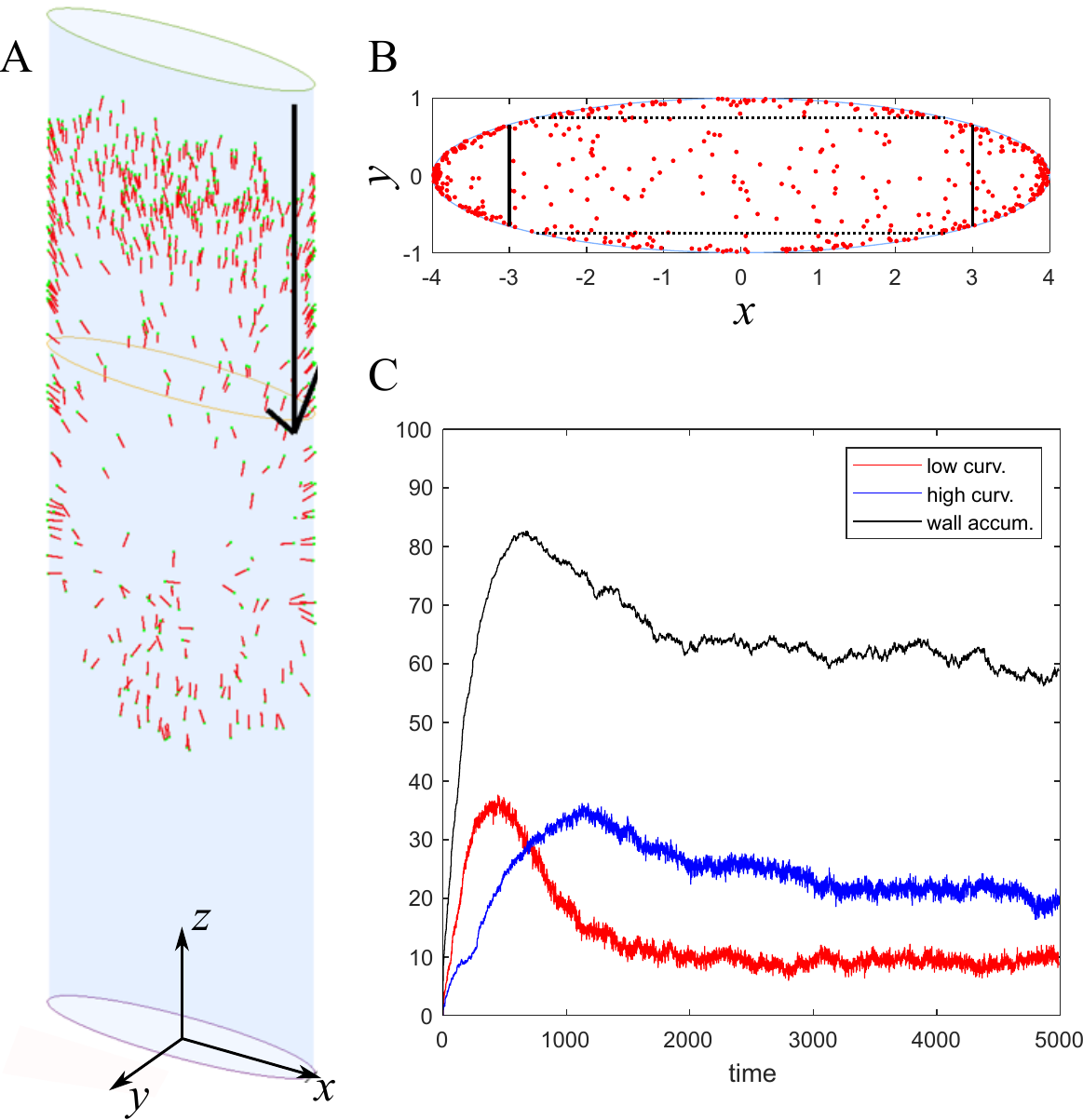}
\caption{{\bf Transport of active rods in an elliptic microchannel with $a=4$ and $b=1$.} Active rods are initiated with $z=0$, random coordinates $x,y$ and orientation angles $\varphi$ and $\theta$. In sub-figure A, $N=500$ active rods are depicted as red segments with the green edge on the active rod's front. The sub-figure shows active rods' configurations after 1000 simulation time step. Black downward arrow shows the direction of the background flow $\boldsymbol{u}$. Purple (bottom), yellow, and green (top) elliptic rings are at $z=-20$, $z=0$, and $z=10$, respectively. Sub-figure B depicts projections of active rods's centers onto $x,y$-plane as thick red dots. Black solid and broken lines are given by $x=\pm n_a$ and $y= \pm n_b$, respectively. Sub-figure (C) depicts fraction of the total number of accumulated rods ({wall accum.}), accumulated rods in $M_a$ (high curv.), and accumulated rods in $M_b$ (low curv.).         }
\label{fig:2}
\end{figure}
\noindent{\it Active rods concentrate at the highest curvature points in the elliptical channel.} To understand how the statistical properties of active rod trajectories change when the ubiquitous geometry of a cylindrical channel with a circular cross-section is perturbed, we consider an elliptical cross-section. That is, we consider the microchannel wall  described as $\partial M = \{(x,y,z)|x^2/a^2+y^2/b^2=1\}$, where $a$ and $b$ are semi-major and semi-minor axes, $a>b$.  The Figure~\ref{fig:2} depicts the results of numerical simulations for $N=500$ rods whose swimming is confined by the microchannel wall $\partial M$.     As evident from this figure (see also {\it SI Video~1}), most of the rods accumulate at the wall and swim upstream, which is the well-known feature of active swimming \cite{potomkin2017focusing}. The background flow vanishes at the wall due to no slip boundary conditions, thus the speed of active rods swimming upstream at the wall is close to the propulsion speed $v_{\text{prop}}$. Active rods swimming downstream near the center-line where the $\boldsymbol{u}$ is maximum have speed $|\boldsymbol{u}|+v_{\text{prop}}$. As the result, right after simulation starts, two distinct profiles of active rods emerge. The first one is the downstream parabolic profile whose spreading speed depends on the flow rate, and this profile quickly dissolves as active rods turn towards the wall. The second one is the upstream flat profile with a constant speed independent from the flow rate. These profiles and how they depend on the flow rate can be observed in {\it SI Video~2}. A typical individual trajectory of an active rod consists of alternating intervals of swimming upstream at the wall and downstream inside the channel. Note that an active rod in our model can escape from the wall into the channel only by random reorientation accounted by the rotational diffusion term in \eqref{subeqn:p}. In the same time, the background flow exerts the maximal torque at the wall where shear rate, $|\nabla \boldsymbol{u}|$, is the highest. This torque is the reason of the upstream swimming (the negative rheotaxis) at the wall but it cannot lead to the escape of the active rod from the wall since this torque vanishes as the active rod becomes tangential to the wall. 

The main result from Figure \ref{fig:2} is that active rods accumulate along lines where the wall curvature is highest, $(\pm a,0,z)$. To measure the ratio between number of rods accumulated at the highest curvature locations $(\pm a,0,z)$ and the lowest curvature locations $(0,\pm b,z)$, we count active rods inside domains $M_a=M\cap \{|x|>n_a\}$ and $M_b=M\cap\{|y|>n_b\}$. $M_a$ and $M_b$ can be understood as vicinities inside the microchannel of locations $(\pm a,0,z)$ and $(0,\pm b, z)$, respectively. Numbers $n_a$ and $n_b$ are chosen in a such way that $M_a$ and $M_b$ have the same area when projected onto $x,y$-cross-section $\Omega$. One can show that this definition of $n_a$ and $n_b$ implies $n_a/n_b = a/b$. 
Figure~\ref{fig:2}(C) shows the count of accumulated rods as a function of time. The fractions of accumulated rods and rods accumulated inside $M_a$ and $M_b$ were stabilized after approximately 1500 time steps indicating that the the rod's distribution converged to a steady state. The fraction of rods accumulated at points of highest curvature, that is, inside $M_a$, is almost twice as large as the fraction of rods at points of lowest curvature, that is, inside $M_b$. It is worth noting that the wall area in $M_b$ is larger than in $M_a$. Indeed, the length of elliptical curves for projections of $M_a$ and $M_b$ in Figure~\ref{fig:2}(B) are approximately 2.52 and 5.32, respectively. Still, there are more accumulated rods at the wall in $M_a$ than in $M_b$, as shown in the graph of Figure~\ref{fig:2}(C). 

\begin{figure}[t!]
\includegraphics[width=0.4\textwidth]{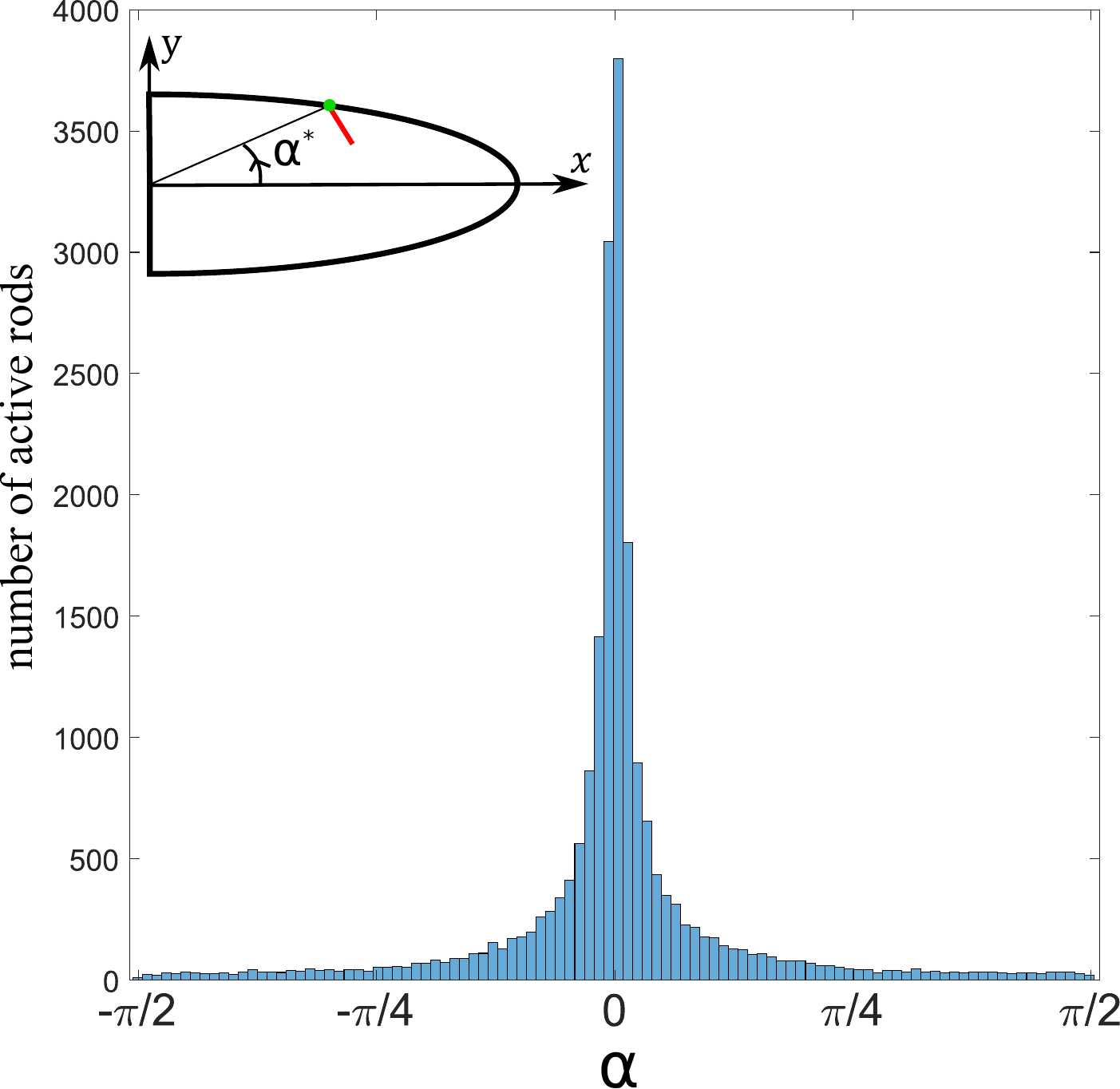}
\caption{{\bf Distribution of wall locations of accumulated active rods with the wall parameterized with the parameter $\alpha$.} $N=50,000$ active rod simulations were used to build this histogram with 100 bins. The inset figure in the top left corner illustrates how $\alpha$ is computed for a given active rod. The active rod is depicted as a red segment with the rod's front edge (a green point) on the wall of the right-hand side of the elliptical microchannel. If the active rod touches the wall on the left-hand side, $\alpha$ is computed by first mirroring the rod with respect to the $y$-axis and then calculating $\alpha$ for the reflected rod. Note $\alpha$ is not exactly the angle with respect to $x$-axis. If we denote the latter angle by $\alpha^*$, then $\alpha^{*}=\arcsin(b\sin(\alpha)/\sqrt{a^2\cos^2\alpha+b^2\sin^2\alpha})$.  Locations of rods are taken from the last time step when the active rod distribution has converged to a steady state.    }
\label{fig:3}
\end{figure}

To illustrate the accumulation of active rods at points of highest curvature, we build a histogram of location of active rods at the wall. Namely, we parameterize the wall cross-section $\partial \Omega$ as $x=a \cos(\alpha), \,y =b\sin(\alpha)$ with the parameter $-\pi\leq \alpha < \pi$. Note that $\alpha = 0,\pi$ and $\alpha = \pm \pi/2$ correspond to points of the highest and lowest curvature points, respectively. For the $i^{\text{th}}$ active rod at the last time step, if the rod is accumulated, then its front is on the wall and its coordinate then can be written as $(a\cos(\hat{\alpha}_i),b\sin(\hat{\alpha}_i))$ for some unique $\hat{\alpha}_i$. Since the distribution of $\hat{\alpha}_i$ is $\pi$ periodic (in the limit $N\to \infty$), we compute $\alpha_i = \text{arcsin}(\sin(\hat{\alpha}_i))$ that allows for reducing the range for $\alpha$ to $(-\pi/2,\pi/2)$.  Distribution of $\alpha_i$ is shown in Figure~\ref{fig:3}. A prominent peak is observed at $\alpha = 0$ showing that the active rods tend to accumulate at the highest curvature points.

\smallskip 
\begin{figure*}
\includegraphics[width=0.325\textwidth]{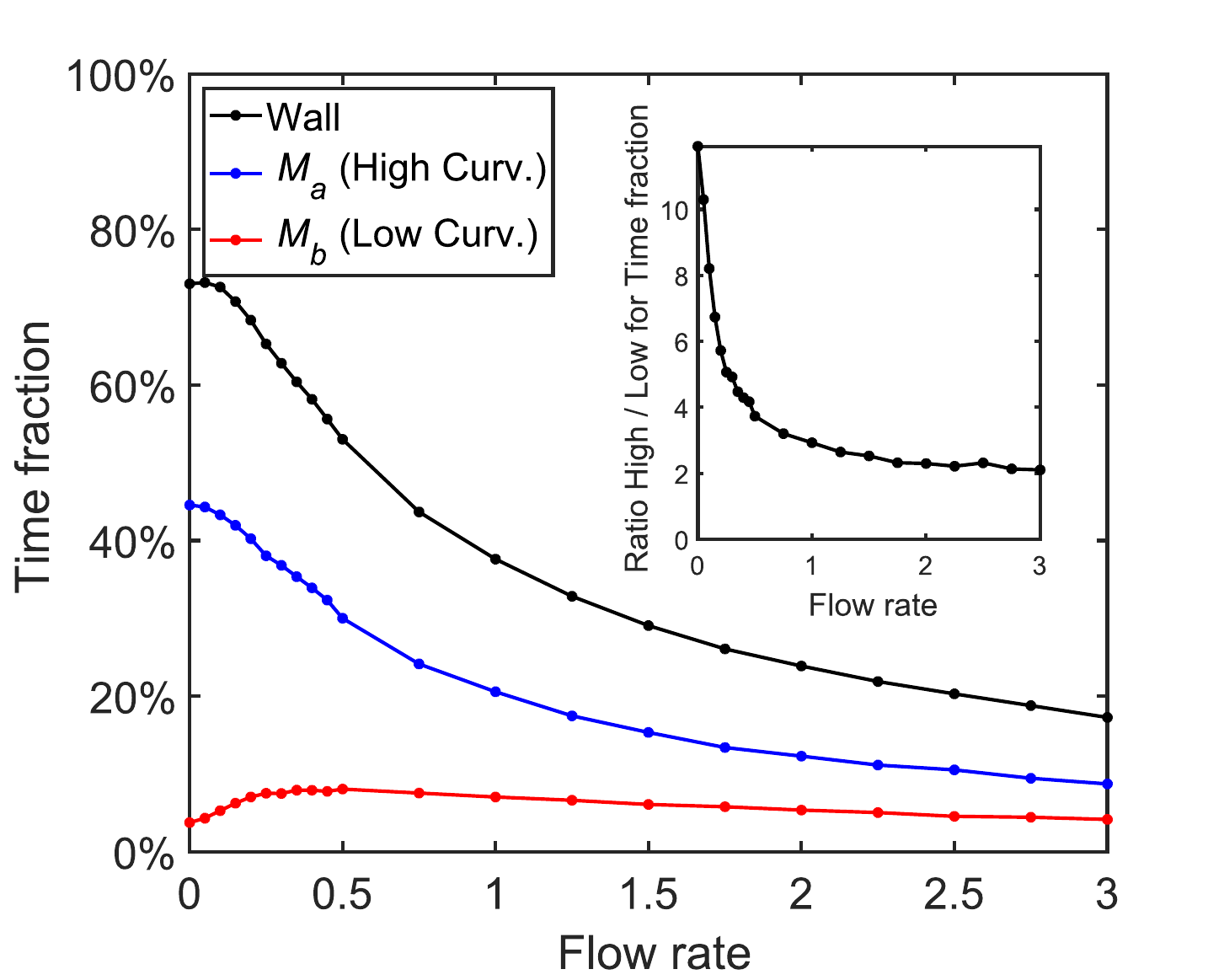}
\includegraphics[width=0.325\textwidth]{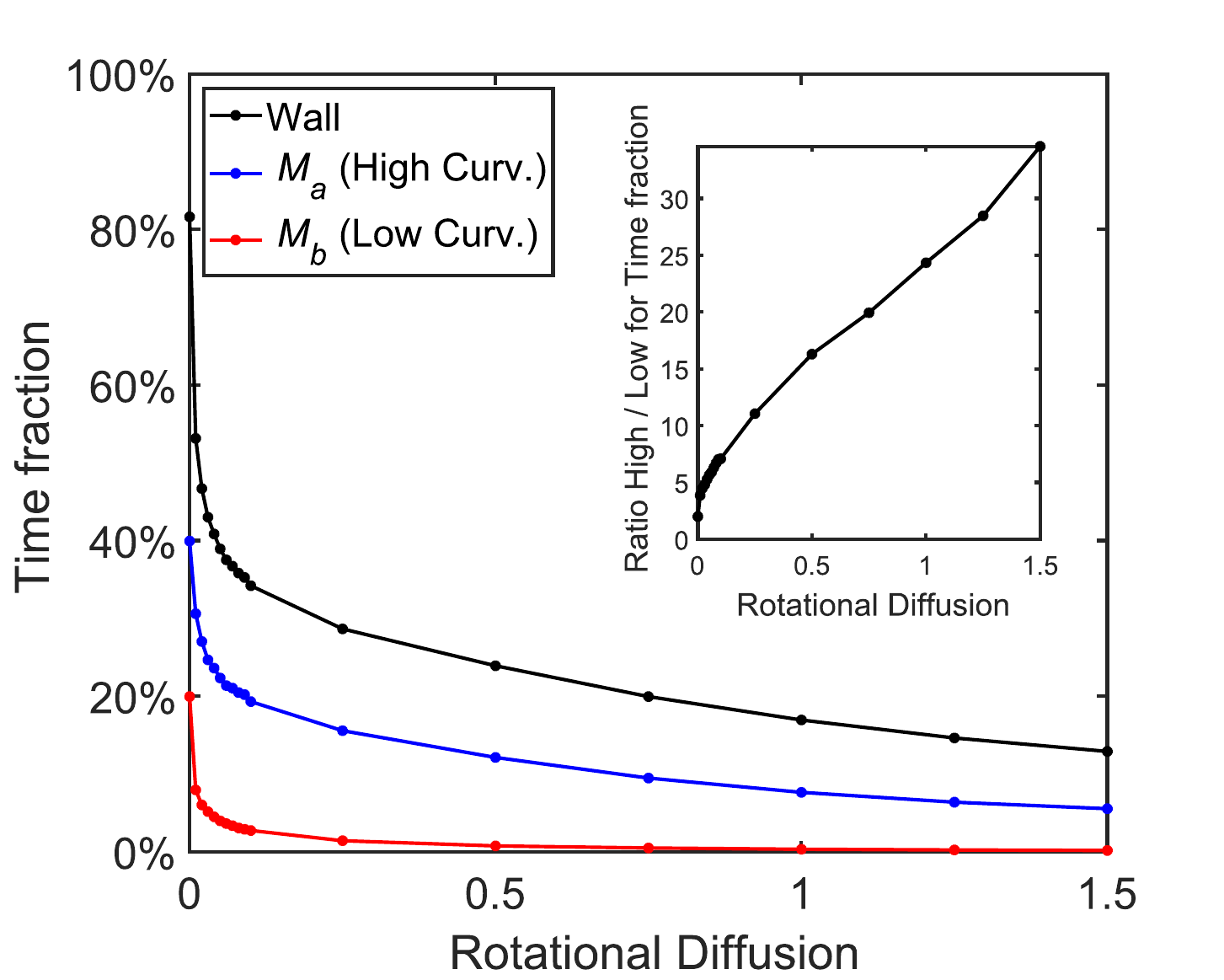}
\includegraphics[width=0.325\textwidth]{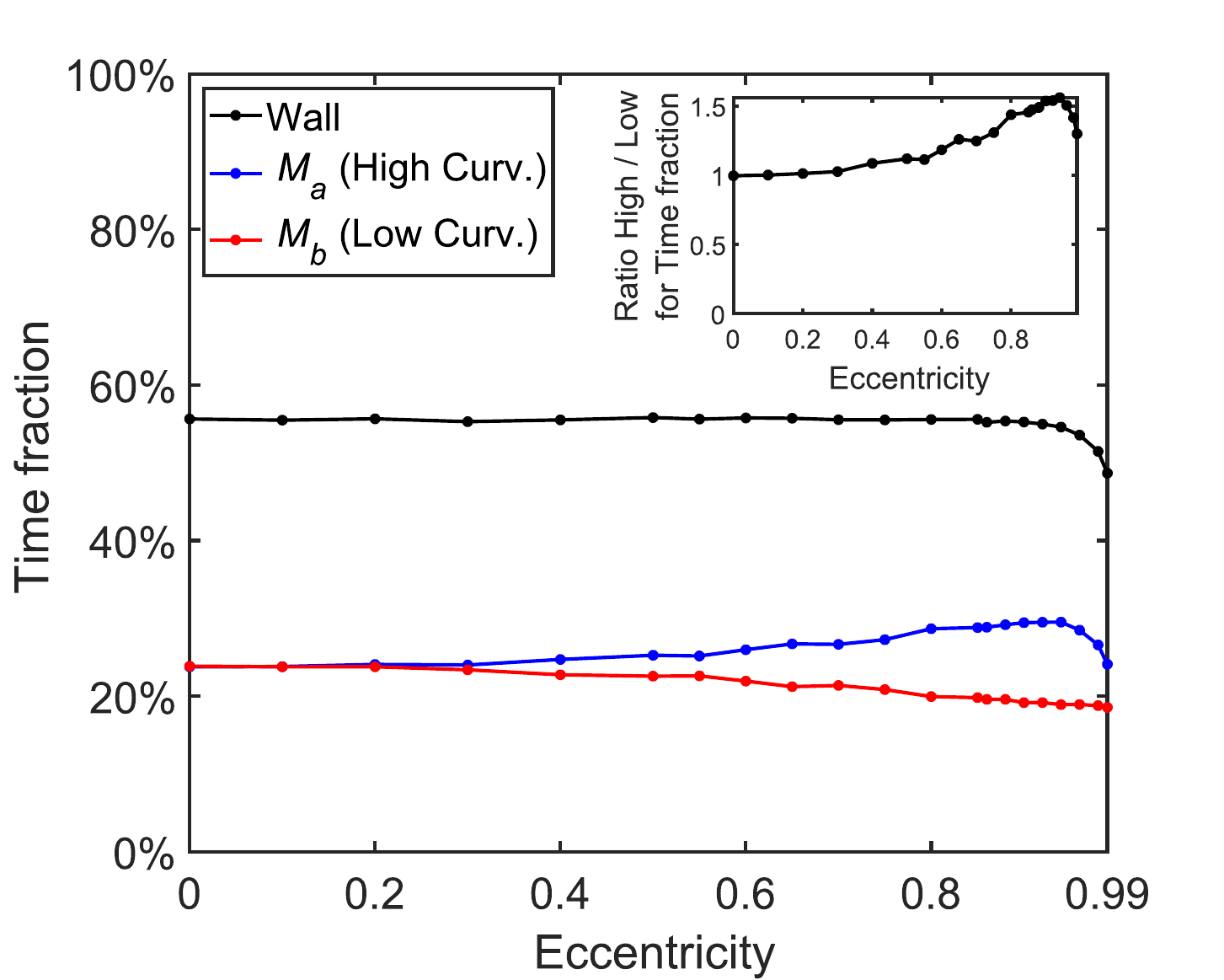}
\caption{{\bf Active rod accumulation dependence on flow rate, rotational diffusion coefficient, and microchannel eccentricity.} Simulations are performed for $R=20,000$ rods. Time fraction in all three sub-figures is the average number of time steps when a rod collides with the wall divided by the total number of time steps. The average is taken over all $R$ rods. The time fraction is expressed in percentages. Typical values of flow rate, diffusion coefficient, and eccentricity are 0.5, 0.01, $\sqrt{15}/4\approx 0.97$, respectively. When varying one of these three parameters, the other two are held at their typical values. The sub-figures also depict time fractions for collisions with the wall inside the domain $M_a$, containing the highest curvature locations (in blue), and $M_b$ with the lowest curvature locations (in red). Insets show the ratio between time fractions for $M_a$ and $M_b$. In the left sub-figure, the flow rate is defined as $|w(0,0)|$ where $w$ is the solution of \eqref{eqn:bvp}. The rotation diffusion in the center sub-figure is the coefficient $\hat{D}_{\text{rot}}$ from \eqref{subeqn:p}. When varying eccentricity (the right sub-figure), we keep $ab$ constant, specifically at $ab=4$. We simulate more values of the flow rate and rotation diffusion coefficient at value $0$, and eccentricity close to $1$ (but excluding 1 since $\Omega$ has no interior in this case) to accurately capture the trend at these parameter limits.    }
\label{fig:4}
\end{figure*}

\noindent{\it Dependence of the active rod accumulation on the flow rate, the diffusion coefficient, and the aspect ratio of the microchannel cross-section.} To quantify the active rod accumulation we calculate  the time fraction spent swimming at the wall, that is, the average percentage of time that an active rod spends at the wall. Note that an equivalent definition of the time fraction is the time average of the wall accumulation depicted as a black graph in Figure~\ref{fig:2}(C).   

We first note that the concentration of active rods at the wall location with the highest curvature is observed without a background flow. However, in the absence of background flow, we no longer observe negative rheotaxis or the asymmetric spreading characterized by an upstream flat and downstream parabolic profile, as there is no upstream or downstream swimming without a background flow.  As shown in Figure~\ref{fig:4}(left), the maximum accumulation occurs near the zero flow rate and  decreases as the flow rate is increased. This decrease is expected: as the flow rate increases, the ratio of self-propulsion to flow rate, $v_{\text{prop}}/|w(0,0)|$, decreases. At high flow rates, active rods behave like passive rods, following streamlines and not accumulating. Therefore, the time fraction will decay to $0$ as the flow rate goes to $\infty$. Nevertheless, for small flow rates we also observe that the average accumulation time does not change significantly. %, and even grows slightly as we increase flow rates $0$ to $0.05$ (specifically, numerical values for these rates are 73.01\% and 73.16\%, respectively). Hence, 
The role of the background flow in active rod accumulation cannot be reduced to merely the decrease of the accumulation. 
% MAYBE USE IN DISCUSSION:
%Indeed, the level of accumulation depends on two factors: the average time it takes for an active rod swimming inside the channel to collide with the wall, $T_{\text{col}}$, and the time it takes for an active rod swimming at the wall to escape from the wall, $T_{\text{esc}}$. If we increase $T_{\text{col}}$ and decrease $T_{\text{esc}}$, which one would expect to happen by increasing the background flow, then the accumulation goes down.      
%%%%%%%% OLD
%We hypothesize that a larger background flow, on the one hand, facilitates a faster collision of an active rod initially swimming inside the macro-channel with the wall, thereby increasing accumulation. On the other hand, a larger background flow reduces the time spent at the wall because a higher shear rate $\nabla \boldsymbol{u}$ at the wall reorients the accumulated active rod more quickly, allowing it to escape sooner and thus decreasing the overall accumulation. We conclude form Figure \ref{fig:4} (left) that the first effect dominates the other for small flow rates, less 0.1, and the second effect is more important that the first one for larger flow rates. 
We observe that the time fraction for accumulation in the low curvature domain $M_b$ grows with the flow rate until the flow rate is approximately $0.5$.
Concentration at the high curvature locations drops down with the flow rate, as it is seen in the inset of Figure~\ref{fig:4}(left).    

Active rod accumulation decreases as the rotational diffusion $\hat{D}_{\text{rot}}$ increases (see Figure~\ref{fig:4}(center)). For zero rotational diffusion ($\hat{D}_{\text{rot}} = 0$), if an active rod collides with the wall, it remains there indefinitely because neither the background flow nor rod-wall interactions can reorient the rod towards the interior of the microchannel. In other words, since all active rods eventually collide with the wall, the time fraction, if computed over a time interval starting long after the initial time step, will be close to 100\% for $\hat{D}_{\text{rot}} = 0$. The observed time fraction for $\hat{D}_{\text{rot}} = 0$ is approximately 83\%, which is notably less than 100\%. This discrepancy occurs because the time fraction is calculated over a period that includes the initial moment when active rods are randomly generated, mostly away from the wall, and thus require some finite time to collide with the wall. As $\hat{D}_{\text{rot}}$ is increased from $0$, accumulation values drop down quickly. Namely, the time fraction halves for $\hat{D}_{\text{rot}}=0.03$ compared to results for $\hat{D}_{\text{rot}}=0$. Interestingly, despite the decrease in the total accumulation as we increase $\hat{D}_{\text{rot}}$, the ratio between accumulation numbers at high and low curvature locations grows, as visible in the inset of Figure~\ref{fig:4}(center).    

Finally, we investigated how the total accumulation and the ratio between accumulations at high and low curvature locations depends on the eccentricity of the microchannel cross-section, $e=\sqrt{1-b^2/a^2}$ (see Figure~\ref{fig:4}(right)). For the sake of comparison, we keep $ab$ constant so that the volume of the fluid passing through a microchannel cross-section per unit of time $\nu_e$ is the same. Keeping $ab$ constant also implies that we consider the same cross-sectional area. Introduce $s=ab$. We find expression of $a$ and $b$ in terms for $s$ and $e$:
\begin{equation*}
a = s^{1/2}(1-e^2)^{-1/4}\quad b = s^{1/2}(1-e^2)^{1/4}.
\end{equation*}
We fix the area parameter $s=4$ and increase eccentricity $e$ from 0 to 0.99. We see in Figure~\ref{fig:4}(right) that the time fraction of active rods accumulating at wall stays constant for a wide range of eccentricities, $0\leq e \lessapprox 0.95$. Time fractions for $M_a$ and $M_b$ are the same if $e=0$, that is $a=b$, which is expected since $e=0$ means that the cross-section is a circle and, thus, $M_a$ and $M_b$ have the same shape. As we increase eccentricity $e$ we see the discrepancy between accumulations at $M_a$ and $M_b$, but this discrepancy decreases as the total accumulation starts decreasing.    
\begin{figure}[t]
\includegraphics[width=0.5\textwidth]{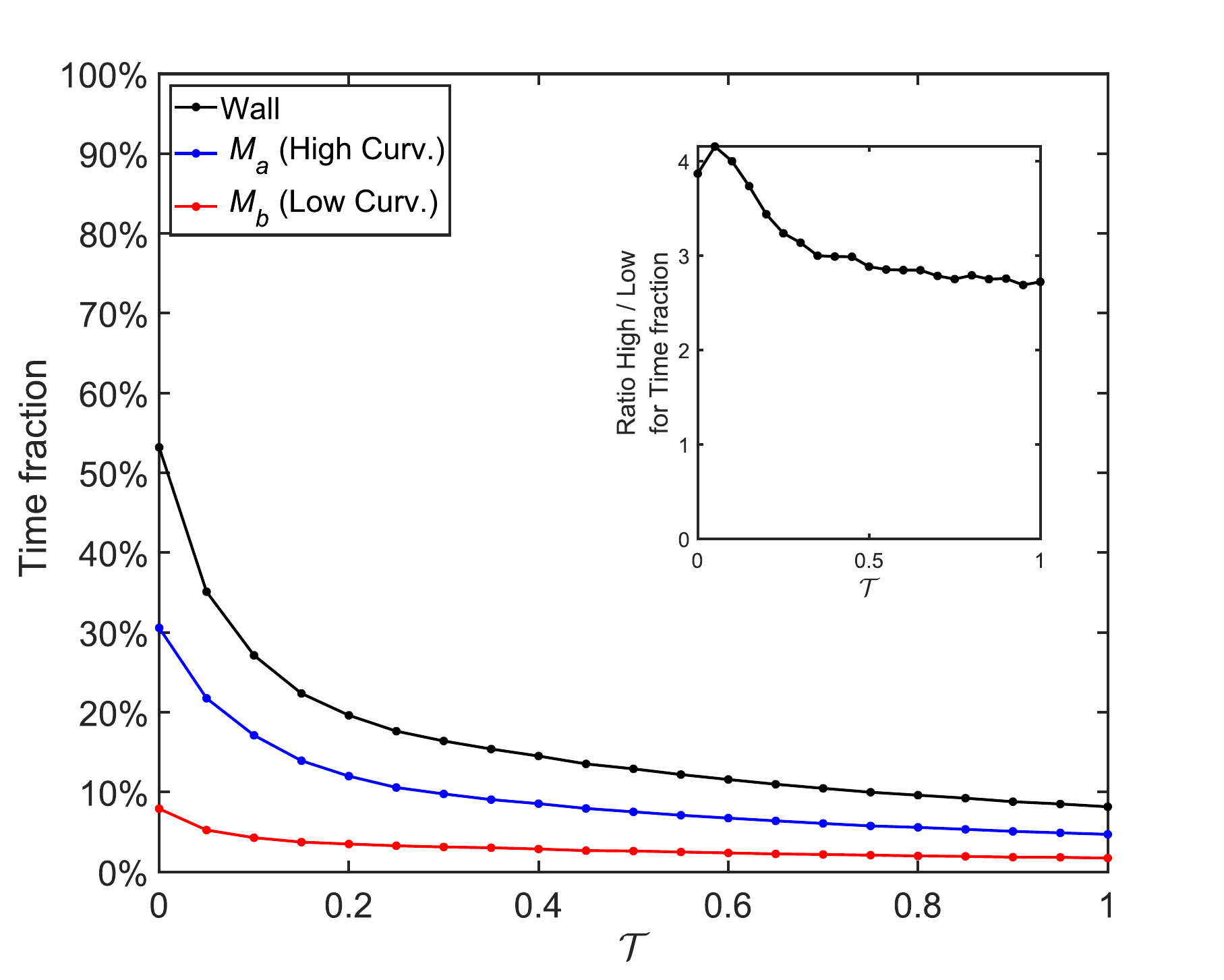}
\caption{{\bf Dependence of active rod accumulation on the wall torque $\mathcal{T}$.} Settings are the same as in plots from Figure~\ref{fig:4}.}
\label{fig:5}
\end{figure}

\noindent{\it Wall torque reduces active rod accumulation.} 
We performed numerical simulations for $0\leq \mathcal{T}\leq 1$ introduced in \eqref{eqn:wall-torque}, with other parameter values held the same as in Figure~\ref{fig:4}. As shown in Figure~\ref{fig:5}, increasing $\mathcal{T}$ results in a decrease of the accumulation of active rods. The wall torque also accelerates the transport of accumulated rods along the wall surface, see also {\it SI Video 3}. This is due to that the wall torque term \eqref{eqn:wall-torque} reorients an accumulated active rod to a direction tangent to the wall surface. When an active rod is oriented tangentially to the wall, it swims faster along the wall, as this speed is proportional to the projection of the direction vector $\boldsymbol{p}$ onto the tangent plane.  In contrast, an active rod that is almost perpendicular to the wall swims more slowly along the wall. Additionally, tangential swimming increases the likelihood of the active rod escaping from the wall due to random reorientation and thus reduces the wall accumulation.   

% (b) Dependence on the flow rate 
% I. Total accumulation
% II. Accumulation at high curvature
% III. Accumulation at low curvature 
% IV. Ratio High / Low
% V. High flow rates - decay rate for accumulation (TO BE DISCUSSED)

% (c) Dependence on diffusion 
% I. Total accumulation
% II. Accumulation at high curvature
% III. Accumulation at low curvature 
% IV. Ratio High / Low
% V. Small D. 

% (d) Dependence on the cross-section aspect ratio 
% I. Total accumulation
% II. Optimal cross-section for given flow rate

% (e) Evidence of steady state
% I. dependence on flow rate and diffusion

% (f) lake problem 
% I. Analytical formula for angular distribution
% II. Discussion of the wall torque: Curvature reduces the aligning effect of the wall torque.

%%%%%%%%%%%%%%%%%%%%%%%%%%%%%%%%%%%%%%%%%%%%%%%%%%%%%%%%%%

%%%%%%%%%%%%%%%%%%%%%%%%%%%%%%%%%%%%%%%%%%%%%%%%%%%%%%%%%%% 2. Rectangles - 4-lane traffic
%%%%%%%%%%%%%%%%%%%%%%%%%%%%%%%%%%%%%%%%%%%%%%%%%%%%%%%%%%

\section{Discussion}
% the lake problem 
To explain why active rods tend to accumulate at the highest curvature points, we consider the problem of active rod swimming in a two-dimensional elliptical lake, that is, we disregard the background flow and dynamics in $z$-dimension. Note that, as observed from Figure~\ref{fig:4}(left), while the background flow does influence the accumulation profiles, it is not the primary cause of the accumulation at the points of highest curvature. We also neglect random reorientation or, in other words, the rotational diffusion. As a consequence, the orientation of the active rod does not change over time, since we disregard all forces that may reorient the rod, such as background shear and rotational diffusion. An active rod will swim straight until it collides with the wall, and then it will slide along the wall until its orientation 
becomes perpendicular to the wall surface, that is,
when the active rod's orientation vector $\boldsymbol{p}$ coincides with the outward normal vector $\boldsymbol{n}$. If we simulate many active rods with random initial locations and orientations, more rods will tend to accumulate at points with the highest curvature. Roughly speaking, this is because vicinities of the highest curvature points have a greater range of direction angles for the outward normal vector compared to vicinities of the same size of points with a lower curvature. 

To make this statement precise, consider an active rod with the initial orientation $\boldsymbol{p} = (\cos(\varphi),\sin(\varphi))^{\text{T}}$ so that the orientation angle $\varphi$ is uniformly random,  $\varphi\sim\mathcal{U}(-\pi,\pi)$. Regardless its initial location, the active rod will converge to the point on the wall where $\boldsymbol{p}=\boldsymbol{n}$ (see also {\it SI Video 4}). The final location of the active rod can be described by $\alpha$, the wall parameter so the wall is parameterized as $x=a \cos (\alpha),\,y=b\sin(\alpha)$ with $-\pi<\alpha\leq\pi$. We show in Supplementary Information that the uniform distributions of the initial orientations $\varphi$ and the condition for the final location $\boldsymbol{p}=\boldsymbol{n}$ imply that the distribution of $\alpha$ is 
\begin{equation}
f(\alpha) = \dfrac{ab}{2\pi(b^2\cos^2(\alpha)+a^2\sin^2(\alpha))}.\label{eqn:distribution}
\end{equation}

\noindent Assuming $a>b$, this probability distribution function $f(\alpha)$, qualitatively similar to the distribution depicted in Figure~\ref{fig:3}, attains the maximum at $\alpha = 0,\pi$, corresponding to the points of the highest curvature, and the minimum at $\alpha = \pm \pi/2$, the lowest curvature points. Nonetheless, the ratio between the maximum and minimum values for the distribution shown in Figure~\ref{fig:3} is much higher than that for $f(\alpha)$. This highlights the influence of the third dimension and rotational diffusion in the simulations presented in Figure~\ref{fig:3}.

% the wall torque 

The lake problem reveals a non-trivial effect of the wall torque on dynamics of active rods accumulated at a curved surface. If the wall is planar, then introduction of the wall torque makes the orientation $\boldsymbol{p}=\boldsymbol{n}$ unstable, regardless of how small the parameter $\mathcal{T}>0$ is. Surprisingly, the orientation $\boldsymbol{p}=\boldsymbol{n}$ can be asymptotically stable for $\mathcal{T}>0$ when the wall has a non-zero curvature. Namely, there is a critical value of the wall torque parameter $\mathcal{T}_{\text{crit}}>0$ at which a pitchfork bifurcation occurs causing active rods to become stuck at the wall with $\boldsymbol{p}=\boldsymbol{n}$ in the sub-critical case $\mathcal{T}<\mathcal{T}_{\text{crit}}$ (see also {\it SI Video~5}). For example, consider an active rod swimming in the circular lake $\Omega$ of radius $R$. Suppose that the active rod has collided with the wall so that its dynamics occurs along the wall. Let $\psi(t)$ denote the angle between the orientation vector $\boldsymbol{p}$ of the active rod and the outward normal vector $\boldsymbol{n}$ at the wall location where the active rod touches the wall at time $t$. If $\psi=0$, then the rod is perpendicular to the wall and neither swims nor reorients. Angles $\psi=\pm \pi/2$ correspond to swimming tangent to the wall. 
We show in Supplementary Information that $\psi(t)$ satisfies the following differential equation: 
\begin{equation}
\dot{\psi}=\dfrac{6v_{\text{prop}}}{\ell}\sin(\psi)\left(\mathcal{T}\cos(\psi)-\dfrac{\ell}{6R}\right).
\end{equation}
If we fix $v_{\text{prop}}$, $\ell$, and $R$, then this differential equation exhibits a pitchfork bifurcation at $\mathcal{T}_{\text{crit}}=\ell/(6R)$. If $0\leq \mathcal{T}\leq \mathcal{T}_{\text{crit}}$, then the active rod eventually accumulates at the wall with $\psi=0$ and no swimming. If $\mathcal{T}>\mathcal{T}_{\text{crit}}$, then $\psi=0$ is unstable and $\psi(t)$ converges to $\pm\arccos(\ell/(6R\mathcal{T}))$ meaning that the active rod will swim along the wall, either clockwise or counterclockwise, at an angle relative to the wall (see also {\it SI Video 6}). If we formally pass to the planar limit $R\to \infty$, then $\mathcal{T}_{\text{crit}}$ vanishes so $\psi=0$ is unstable for all $\mathcal{T}>0$ and the two stable angles are $\pm \arccos(0) = \pm \pi/2$, that is, swimming tangent to the wall. Finally, we note that this bifurcation phenomenon can lead to formation of trapping regions around high curvature points in the elliptical domain $\Omega$. Specifically, because the value of $R$ is inversely proportional to the curvature, the critical value $\mathcal{T}_{\text{crit}}$ increases with the curvature. If  $\mathcal{T}$ is fixed, then in an elliptical wall where curvature varies, it may happen that regions with higher curvature have $\mathcal{T}<\mathcal{T}_{\text{crit}}$ causing active rods to stop swimming with $\psi=0$ (forming the trapping region), whereas locations with lower curvature have $\mathcal{T}>\mathcal{T}_{\text{crit}}$ allowing active rods to maintain a certain speed along the wall and eventually clear the low-curvature regions. An example of such a case is presented in {\it SI Video 5}.  

\smallskip

\noindent{\it Future directions.} 
Our results demonstrate that the wall curvature has a trapping effect on the dynamics of individual active rods, which can be utilized to control and guide micro-swimmers. We hypothesize that corners act as stronger attractors for active rod concentration compared to locations with finite curvature. Specifically, our model captures the four-lane structure of active swimmer dynamics in microchannels with a rectangular cross-section, see {\it SI Video~7}. This four-lane structure indicates that active swimmers tend to move along the four corners of the microchannel, a phenomenon observed in experiments \cite{HilKalMcMKos2007,nuris2015,Bukatin2020}. The accumulation of active rods may also depend on the angle of the corner, which can be tested using the model presented in this work.

To draw strong conclusions about active rod accumulation for different geometries, a kinetic approach is required. This involves solving a partial differential equation where the unknown is the distribution function of the active rods' location and orientation. Unlike Monte Carlo simulations, this approach is deterministic. Such an approach, which also explicitly distinguishes between the distributions of active rods at the wall and away from it, has been developed in \cite{potomkin2020confined} for the two-dimensional case (where the wall is a curve). Implementing this approach for a three-dimensional cylindrical microchannel presents challenges due to the need to handle differential operators over surfaces (the unit sphere for orientation $\boldsymbol{p}$ and the wall surface) and singular terms that describe active rods detaching from the wall. Developing a robust numerical scheme for this equation will enable testing and comparison of various microchannel geometries and help elucidate the effects of curvature and corners on different types of micro-swimmers.

% High curvature wells may guide control transport
% 

% dependence on shear rate 

% curvature vs corners 

% the importance of implementing boundary conditions and kinetic approaches   

% future direction: kinetic approach

% CURVATURE TRAPPING

% ADDITIONAL THINGS TO ADDRESS IF TIME PERMITS 

% 0. FLOW BOTH HELPS AND HINDERS ACCUMULATION: COMPUTE ACCUMULATION TIMES AND ESCAPE TIMES. 
% 1. STEADY-STATE TIME VS FLOW RATE
% 2. ACCUMULATION AS FLOW RATE \to \infty
% 3. ACCUMULATION WITH WALL TORQUE

% OLD plan 
% 1. Robustness of ellipsoid results
% 2. Effect of corners (periodic boundary conditions)
% 3. Trapezoid calculations

\bigskip 

\begin{acknowledgments}
The work of M.P. was supported by the Hellman Fellowship Program 2024-2025. 
\end{acknowledgments}

%\appendix

%\section{Appendixes}

% The \nocite command causes all entries in a bibliography to be printed out
% whether or not they are actually referenced in the text. This is appropriate
% for the sample file to show the different styles of references, but authors
% most likely will not want to use it.
%\nocite{*}

\bibliography{refs}% Produces the bibliography via BibTeX.

%apsrev4-2.bst 2019-01-14 (MD) hand-edited version of apsrev4-1.bst
%Control: key (0)
%Control: author (8) initials jnrlst
%Control: editor formatted (1) identically to author
%Control: production of article title (0) allowed
%Control: page (0) single
%Control: year (1) truncated
%Control: production of eprint (0) enabled
\begin{thebibliography}{37}%
\makeatletter
\providecommand \@ifxundefined [1]{%
 \@ifx{#1\undefined}
}%
\providecommand \@ifnum [1]{%
 \ifnum #1\expandafter \@firstoftwo
 \else \expandafter \@secondoftwo
 \fi
}%
\providecommand \@ifx [1]{%
 \ifx #1\expandafter \@firstoftwo
 \else \expandafter \@secondoftwo
 \fi
}%
\providecommand \natexlab [1]{#1}%
\providecommand \enquote  [1]{``#1''}%
\providecommand \bibnamefont  [1]{#1}%
\providecommand \bibfnamefont [1]{#1}%
\providecommand \citenamefont [1]{#1}%
\providecommand \href@noop [0]{\@secondoftwo}%
\providecommand \href [0]{\begingroup \@sanitize@url \@href}%
\providecommand \@href[1]{\@@startlink{#1}\@@href}%
\providecommand \@@href[1]{\endgroup#1\@@endlink}%
\providecommand \@sanitize@url [0]{\catcode `\\12\catcode `\$12\catcode
  `\&12\catcode `\#12\catcode `\^12\catcode `\_12\catcode `\%12\relax}%
\providecommand \@@startlink[1]{}%
\providecommand \@@endlink[0]{}%
\providecommand \url  [0]{\begingroup\@sanitize@url \@url }%
\providecommand \@url [1]{\endgroup\@href {#1}{\urlprefix }}%
\providecommand \urlprefix  [0]{URL }%
\providecommand \Eprint [0]{\href }%
\providecommand \doibase [0]{https://doi.org/}%
\providecommand \selectlanguage [0]{\@gobble}%
\providecommand \bibinfo  [0]{\@secondoftwo}%
\providecommand \bibfield  [0]{\@secondoftwo}%
\providecommand \translation [1]{[#1]}%
\providecommand \BibitemOpen [0]{}%
\providecommand \bibitemStop [0]{}%
\providecommand \bibitemNoStop [0]{.\EOS\space}%
\providecommand \EOS [0]{\spacefactor3000\relax}%
\providecommand \BibitemShut  [1]{\csname bibitem#1\endcsname}%
\let\auto@bib@innerbib\@empty
%</preamble>
\bibitem [{\citenamefont {Aranson}(2022)}]{aranson2022bacterial}%
  \BibitemOpen
  \bibfield  {author} {\bibinfo {author} {\bibfnamefont {I.~S.}\ \bibnamefont
  {Aranson}},\ }\bibfield  {title} {\bibinfo {title} {Bacterial active
  matter},\ }\href@noop {} {\bibfield  {journal} {\bibinfo  {journal} {Reports
  on Progress in Physics}\ }\textbf {\bibinfo {volume} {85}},\ \bibinfo {pages}
  {076601} (\bibinfo {year} {2022})}\BibitemShut {NoStop}%
\bibitem [{\citenamefont {Paxton}\ \emph {et~al.}(2005)\citenamefont {Paxton},
  \citenamefont {Sen},\ and\ \citenamefont {Mallouk}}]{PaxSenMal05}%
  \BibitemOpen
  \bibfield  {author} {\bibinfo {author} {\bibfnamefont {W.}~\bibnamefont
  {Paxton}}, \bibinfo {author} {\bibfnamefont {A.}~\bibnamefont {Sen}},\ and\
  \bibinfo {author} {\bibfnamefont {T.}~\bibnamefont {Mallouk}},\ }\bibfield
  {title} {\bibinfo {title} {{M}otility of {C}atalytic {N}anoparticles
  {T}hrough {S}elf-{G}enerated {F}orces},\ }\href@noop {} {\bibfield  {journal}
  {\bibinfo  {journal} {Chem. Eur. J.}\ }\textbf {\bibinfo {volume} {11}},\
  \bibinfo {pages} {6462} (\bibinfo {year} {2005})}\BibitemShut {NoStop}%
\bibitem [{\citenamefont {Paxton}\ \emph
  {et~al.}(2006{\natexlab{a}})\citenamefont {Paxton}, \citenamefont
  {Sundararajan}, \citenamefont {Mallouk},\ and\ \citenamefont
  {Sen}}]{PaxSunMalSen06}%
  \BibitemOpen
  \bibfield  {author} {\bibinfo {author} {\bibfnamefont {W.}~\bibnamefont
  {Paxton}}, \bibinfo {author} {\bibfnamefont {S.}~\bibnamefont
  {Sundararajan}}, \bibinfo {author} {\bibfnamefont {T.}~\bibnamefont
  {Mallouk}},\ and\ \bibinfo {author} {\bibfnamefont {A.}~\bibnamefont {Sen}},\
  }\bibfield  {title} {\bibinfo {title} {Chemical {L}ocomotion},\ }\href@noop
  {} {\bibfield  {journal} {\bibinfo  {journal} {Angew. Chem. Int. Ed. Engl.}\
  }\textbf {\bibinfo {volume} {45}},\ \bibinfo {pages} {5420} (\bibinfo {year}
  {2006}{\natexlab{a}})}\BibitemShut {NoStop}%
\bibitem [{\citenamefont {Paxton}\ \emph
  {et~al.}(2006{\natexlab{b}})\citenamefont {Paxton}, \citenamefont {Baker},
  \citenamefont {Kline}, \citenamefont {Wang}, \citenamefont {Mallouk},\ and\
  \citenamefont {Sen}}]{PaxBakKliWanMalSen06}%
  \BibitemOpen
  \bibfield  {author} {\bibinfo {author} {\bibfnamefont {W.}~\bibnamefont
  {Paxton}}, \bibinfo {author} {\bibfnamefont {P.}~\bibnamefont {Baker}},
  \bibinfo {author} {\bibfnamefont {T.}~\bibnamefont {Kline}}, \bibinfo
  {author} {\bibfnamefont {Y.}~\bibnamefont {Wang}}, \bibinfo {author}
  {\bibfnamefont {T.}~\bibnamefont {Mallouk}},\ and\ \bibinfo {author}
  {\bibfnamefont {A.}~\bibnamefont {Sen}},\ }\bibfield  {title} {\bibinfo
  {title} {Catalytically {I}nduced {E}lectrokinetics for {M}otors and
  {M}icropumps},\ }\href@noop {} {\bibfield  {journal} {\bibinfo  {journal} {J.
  Am. Chem. Soc.}\ }\textbf {\bibinfo {volume} {128}},\ \bibinfo {pages}
  {14881} (\bibinfo {year} {2006}{\natexlab{b}})}\BibitemShut {NoStop}%
\bibitem [{\citenamefont {Wensink}\ \emph {et~al.}(2012)\citenamefont
  {Wensink}, \citenamefont {Dunkel}, \citenamefont {Heidenreich}, \citenamefont
  {Drescher}, \citenamefont {Goldstein}, \citenamefont {L{\"o}wen},\ and\
  \citenamefont {Yeomans}}]{wensink2012meso}%
  \BibitemOpen
  \bibfield  {author} {\bibinfo {author} {\bibfnamefont {H.}~\bibnamefont
  {Wensink}}, \bibinfo {author} {\bibfnamefont {J.}~\bibnamefont {Dunkel}},
  \bibinfo {author} {\bibfnamefont {S.}~\bibnamefont {Heidenreich}}, \bibinfo
  {author} {\bibfnamefont {K.}~\bibnamefont {Drescher}}, \bibinfo {author}
  {\bibfnamefont {R.}~\bibnamefont {Goldstein}}, \bibinfo {author}
  {\bibfnamefont {H.}~\bibnamefont {L{\"o}wen}},\ and\ \bibinfo {author}
  {\bibfnamefont {J.~M.}\ \bibnamefont {Yeomans}},\ }\bibfield  {title}
  {\bibinfo {title} {Meso-scale turbulence in living fluids},\ }\href@noop {}
  {\bibfield  {journal} {\bibinfo  {journal} {Proceedings of the National
  Academy of Sciences}\ }\textbf {\bibinfo {volume} {109}},\ \bibinfo {pages}
  {14308} (\bibinfo {year} {2012})}\BibitemShut {NoStop}%
\bibitem [{\citenamefont {Sokolov}\ and\ \citenamefont
  {Aranson}(2012)}]{SokAra2012}%
  \BibitemOpen
  \bibfield  {author} {\bibinfo {author} {\bibfnamefont {A.}~\bibnamefont
  {Sokolov}}\ and\ \bibinfo {author} {\bibfnamefont {I.}~\bibnamefont
  {Aranson}},\ }\bibfield  {title} {\bibinfo {title} {Physical properties of
  collective motion in suspensions of bacteria},\ }\href@noop {} {\bibfield
  {journal} {\bibinfo  {journal} {Physical Review Letters}\ }\textbf {\bibinfo
  {volume} {109}},\ \bibinfo {pages} {248109} (\bibinfo {year}
  {2012})}\BibitemShut {NoStop}%
\bibitem [{\citenamefont {Sokolov}\ \emph {et~al.}(2009)\citenamefont
  {Sokolov}, \citenamefont {Goldstein}, \citenamefont {Feldchtein},\ and\
  \citenamefont {Aranson}}]{SokGolFelAra2009}%
  \BibitemOpen
  \bibfield  {author} {\bibinfo {author} {\bibfnamefont {A.}~\bibnamefont
  {Sokolov}}, \bibinfo {author} {\bibfnamefont {R.}~\bibnamefont {Goldstein}},
  \bibinfo {author} {\bibfnamefont {F.}~\bibnamefont {Feldchtein}},\ and\
  \bibinfo {author} {\bibfnamefont {I.}~\bibnamefont {Aranson}},\ }\bibfield
  {title} {\bibinfo {title} {Enhanced mixing and spatial instability in
  concentrated bacterial suspensions},\ }\href@noop {} {\bibfield  {journal}
  {\bibinfo  {journal} {Physical Review E}\ }\textbf {\bibinfo {volume} {80}},\
  \bibinfo {pages} {031903} (\bibinfo {year} {2009})}\BibitemShut {NoStop}%
\bibitem [{\citenamefont {Gompper}\ \emph {et~al.}(2020)\citenamefont
  {Gompper}, \citenamefont {Winkler}, \citenamefont {Speck}, \citenamefont
  {Solon}, \citenamefont {Nardini}, \citenamefont {Peruani}, \citenamefont
  {L{\"o}wen}, \citenamefont {Golestanian}, \citenamefont {Kaupp},
  \citenamefont {Alvarez} \emph {et~al.}}]{gompper20202020}%
  \BibitemOpen
  \bibfield  {author} {\bibinfo {author} {\bibfnamefont {G.}~\bibnamefont
  {Gompper}}, \bibinfo {author} {\bibfnamefont {R.~G.}\ \bibnamefont
  {Winkler}}, \bibinfo {author} {\bibfnamefont {T.}~\bibnamefont {Speck}},
  \bibinfo {author} {\bibfnamefont {A.}~\bibnamefont {Solon}}, \bibinfo
  {author} {\bibfnamefont {C.}~\bibnamefont {Nardini}}, \bibinfo {author}
  {\bibfnamefont {F.}~\bibnamefont {Peruani}}, \bibinfo {author} {\bibfnamefont
  {H.}~\bibnamefont {L{\"o}wen}}, \bibinfo {author} {\bibfnamefont
  {R.}~\bibnamefont {Golestanian}}, \bibinfo {author} {\bibfnamefont {U.~B.}\
  \bibnamefont {Kaupp}}, \bibinfo {author} {\bibfnamefont {L.}~\bibnamefont
  {Alvarez}}, \emph {et~al.},\ }\bibfield  {title} {\bibinfo {title} {The 2020
  motile active matter roadmap},\ }\href@noop {} {\bibfield  {journal}
  {\bibinfo  {journal} {Journal of Physics: Condensed Matter}\ }\textbf
  {\bibinfo {volume} {32}},\ \bibinfo {pages} {193001} (\bibinfo {year}
  {2020})}\BibitemShut {NoStop}%
\bibitem [{\citenamefont {B\"{a}r}\ \emph {et~al.}(2020)\citenamefont
  {B\"{a}r}, \citenamefont {Gro{\ss}mann}, \citenamefont {Heidenreich},\ and\
  \citenamefont {Peruani}}]{peruani2020review}%
  \BibitemOpen
  \bibfield  {author} {\bibinfo {author} {\bibfnamefont {M.}~\bibnamefont
  {B\"{a}r}}, \bibinfo {author} {\bibfnamefont {R.}~\bibnamefont
  {Gro{\ss}mann}}, \bibinfo {author} {\bibfnamefont {S.}~\bibnamefont
  {Heidenreich}},\ and\ \bibinfo {author} {\bibfnamefont {F.}~\bibnamefont
  {Peruani}},\ }\bibfield  {title} {\bibinfo {title} {Self-propelled rods:
  {I}nsights and perspectives for active matter},\ }\href@noop {} {\bibfield
  {journal} {\bibinfo  {journal} {Annual Review of Condensed Matter Physics}\
  }\textbf {\bibinfo {volume} {11}},\ \bibinfo {pages} {441} (\bibinfo {year}
  {2020})}\BibitemShut {NoStop}%
\bibitem [{\citenamefont {Hallatschek}\ \emph {et~al.}(2023)\citenamefont
  {Hallatschek}, \citenamefont {Datta}, \citenamefont {Drescher}, \citenamefont
  {Dunkel}, \citenamefont {Elgeti}, \citenamefont {Waclaw},\ and\ \citenamefont
  {Wingreen}}]{hallatschek2023proliferating}%
  \BibitemOpen
  \bibfield  {author} {\bibinfo {author} {\bibfnamefont {O.}~\bibnamefont
  {Hallatschek}}, \bibinfo {author} {\bibfnamefont {S.~S.}\ \bibnamefont
  {Datta}}, \bibinfo {author} {\bibfnamefont {K.}~\bibnamefont {Drescher}},
  \bibinfo {author} {\bibfnamefont {J.}~\bibnamefont {Dunkel}}, \bibinfo
  {author} {\bibfnamefont {J.}~\bibnamefont {Elgeti}}, \bibinfo {author}
  {\bibfnamefont {B.}~\bibnamefont {Waclaw}},\ and\ \bibinfo {author}
  {\bibfnamefont {N.~S.}\ \bibnamefont {Wingreen}},\ }\bibfield  {title}
  {\bibinfo {title} {Proliferating active matter},\ }\href@noop {} {\bibfield
  {journal} {\bibinfo  {journal} {Nature Reviews Physics}\ }\textbf {\bibinfo
  {volume} {5}},\ \bibinfo {pages} {407} (\bibinfo {year} {2023})}\BibitemShut
  {NoStop}%
\bibitem [{\citenamefont {Sokolov}\ and\ \citenamefont
  {Aranson}(2009)}]{sokolov2009reduction}%
  \BibitemOpen
  \bibfield  {author} {\bibinfo {author} {\bibfnamefont {A.}~\bibnamefont
  {Sokolov}}\ and\ \bibinfo {author} {\bibfnamefont {I.}~\bibnamefont
  {Aranson}},\ }\bibfield  {title} {\bibinfo {title} {Reduction of viscosity in
  suspension of swimming bacteria},\ }\href@noop {} {\bibfield  {journal}
  {\bibinfo  {journal} {Physical Review Letters}\ }\textbf {\bibinfo {volume}
  {103}},\ \bibinfo {pages} {148101} (\bibinfo {year} {2009})}\BibitemShut
  {NoStop}%
\bibitem [{\citenamefont {Haines}\ \emph {et~al.}(2009)\citenamefont {Haines},
  \citenamefont {Sokolov}, \citenamefont {Aranson}, \citenamefont {Berlyand},\
  and\ \citenamefont {Karpeev}}]{HaiSokAraBer09}%
  \BibitemOpen
  \bibfield  {author} {\bibinfo {author} {\bibfnamefont {B.~M.}\ \bibnamefont
  {Haines}}, \bibinfo {author} {\bibfnamefont {A.}~\bibnamefont {Sokolov}},
  \bibinfo {author} {\bibfnamefont {I.~S.}\ \bibnamefont {Aranson}}, \bibinfo
  {author} {\bibfnamefont {L.}~\bibnamefont {Berlyand}},\ and\ \bibinfo
  {author} {\bibfnamefont {D.~A.}\ \bibnamefont {Karpeev}},\ }\bibfield
  {title} {\bibinfo {title} {A three-dimensional model for the effective
  viscosity of bacterial suspensions},\ }\href@noop {} {\bibfield  {journal}
  {\bibinfo  {journal} {Phys. Rev. E}\ }\textbf {\bibinfo {volume} {80}},\
  \bibinfo {pages} {041922} (\bibinfo {year} {2009})}\BibitemShut {NoStop}%
\bibitem [{\citenamefont {Ryan}\ \emph {et~al.}(2011)\citenamefont {Ryan},
  \citenamefont {Haines}, \citenamefont {Berlyand}, \citenamefont {Ziebert},\
  and\ \citenamefont {Aranson}}]{RyaHaiBerZieAra11}%
  \BibitemOpen
  \bibfield  {author} {\bibinfo {author} {\bibfnamefont {S.}~\bibnamefont
  {Ryan}}, \bibinfo {author} {\bibfnamefont {B.}~\bibnamefont {Haines}},
  \bibinfo {author} {\bibfnamefont {L.}~\bibnamefont {Berlyand}}, \bibinfo
  {author} {\bibfnamefont {F.}~\bibnamefont {Ziebert}},\ and\ \bibinfo {author}
  {\bibfnamefont {I.}~\bibnamefont {Aranson}},\ }\bibfield  {title} {\bibinfo
  {title} {Viscosity of bacterial suspensions: {H}ydrodynamic interactions and
  self-induced noise},\ }\href@noop {} {\bibfield  {journal} {\bibinfo
  {journal} {Rapid Communication to Phys. Rev. E}\ }\textbf {\bibinfo {volume}
  {83}},\ \bibinfo {pages} {050904} (\bibinfo {year} {2011})}\BibitemShut
  {NoStop}%
\bibitem [{\citenamefont {Potomkin}\ \emph {et~al.}(2016)\citenamefont
  {Potomkin}, \citenamefont {Ryan},\ and\ \citenamefont
  {Berlyand}}]{potomkin2016ev}%
  \BibitemOpen
  \bibfield  {author} {\bibinfo {author} {\bibfnamefont {M.}~\bibnamefont
  {Potomkin}}, \bibinfo {author} {\bibfnamefont {S.}~\bibnamefont {Ryan}},\
  and\ \bibinfo {author} {\bibfnamefont {L.}~\bibnamefont {Berlyand}},\
  }\bibfield  {title} {\bibinfo {title} {Effective rheological properties in
  semi-dilute bacterial suspensions},\ }\href@noop {} {\bibfield  {journal}
  {\bibinfo  {journal} {Bulletin of mathematical biology}\ }\textbf {\bibinfo
  {volume} {78}},\ \bibinfo {pages} {580} (\bibinfo {year} {2016})}\BibitemShut
  {NoStop}%
\bibitem [{\citenamefont {Hill}\ \emph {et~al.}(2007)\citenamefont {Hill},
  \citenamefont {Kalkanci}, \citenamefont {McMurry},\ and\ \citenamefont
  {Koser}}]{HilKalMcMKos2007}%
  \BibitemOpen
  \bibfield  {author} {\bibinfo {author} {\bibfnamefont {J.}~\bibnamefont
  {Hill}}, \bibinfo {author} {\bibfnamefont {O.}~\bibnamefont {Kalkanci}},
  \bibinfo {author} {\bibfnamefont {J.}~\bibnamefont {McMurry}},\ and\ \bibinfo
  {author} {\bibfnamefont {H.}~\bibnamefont {Koser}},\ }\bibfield  {title}
  {\bibinfo {title} {Hydrodynamic surface interactions enable escherichia coli
  to seek efficient routes to swim upstream},\ }\href@noop {} {\bibfield
  {journal} {\bibinfo  {journal} {Physical review letters}\ }\textbf {\bibinfo
  {volume} {98}},\ \bibinfo {pages} {068101} (\bibinfo {year}
  {2007})}\BibitemShut {NoStop}%
\bibitem [{\citenamefont {Fu}\ \emph {et~al.}(2012)\citenamefont {Fu},
  \citenamefont {Powers},\ and\ \citenamefont {Stocker}}]{fu2012bacterial}%
  \BibitemOpen
  \bibfield  {author} {\bibinfo {author} {\bibfnamefont {H.~C.}\ \bibnamefont
  {Fu}}, \bibinfo {author} {\bibfnamefont {T.~R.}\ \bibnamefont {Powers}},\
  and\ \bibinfo {author} {\bibfnamefont {R.}~\bibnamefont {Stocker}},\
  }\bibfield  {title} {\bibinfo {title} {Bacterial rheotaxis},\ }\href@noop {}
  {\bibfield  {journal} {\bibinfo  {journal} {Proceedings of the National
  Academy of Sciences}\ }\textbf {\bibinfo {volume} {109}},\ \bibinfo {pages}
  {4780} (\bibinfo {year} {2012})}\BibitemShut {NoStop}%
\bibitem [{\citenamefont {Palacci}\ \emph {et~al.}(2015)\citenamefont
  {Palacci}, \citenamefont {Sacanna}, \citenamefont {Abramian}, \citenamefont
  {Barral}, \citenamefont {Hanson}, \citenamefont {Grosberg}, \citenamefont
  {Pine},\ and\ \citenamefont {Chaikin}}]{PalSacAbrBarHanGroPinCha2015}%
  \BibitemOpen
  \bibfield  {author} {\bibinfo {author} {\bibfnamefont {J.}~\bibnamefont
  {Palacci}}, \bibinfo {author} {\bibfnamefont {S.}~\bibnamefont {Sacanna}},
  \bibinfo {author} {\bibfnamefont {A.}~\bibnamefont {Abramian}}, \bibinfo
  {author} {\bibfnamefont {J.}~\bibnamefont {Barral}}, \bibinfo {author}
  {\bibfnamefont {K.}~\bibnamefont {Hanson}}, \bibinfo {author} {\bibfnamefont
  {A.}~\bibnamefont {Grosberg}}, \bibinfo {author} {\bibfnamefont
  {D.}~\bibnamefont {Pine}},\ and\ \bibinfo {author} {\bibfnamefont
  {P.}~\bibnamefont {Chaikin}},\ }\bibfield  {title} {\bibinfo {title}
  {Artificial rheotaxis},\ }\href@noop {} {\bibfield  {journal} {\bibinfo
  {journal} {Science Advances}\ }\textbf {\bibinfo {volume} {1}},\ \bibinfo
  {pages} {e1400214} (\bibinfo {year} {2015})}\BibitemShut {NoStop}%
\bibitem [{\citenamefont {Yuan}\ \emph {et~al.}(2015)\citenamefont {Yuan},
  \citenamefont {Raizen},\ and\ \citenamefont {Bau}}]{YuaRaiBau2015}%
  \BibitemOpen
  \bibfield  {author} {\bibinfo {author} {\bibfnamefont {J.}~\bibnamefont
  {Yuan}}, \bibinfo {author} {\bibfnamefont {D.}~\bibnamefont {Raizen}},\ and\
  \bibinfo {author} {\bibfnamefont {H.}~\bibnamefont {Bau}},\ }\bibfield
  {title} {\bibinfo {title} {A hydrodynamic mechanism for attraction of
  undulatory microswimmers to surfaces (bordertaxis)},\ }\href@noop {}
  {\bibfield  {journal} {\bibinfo  {journal} {Journal of the Royal Society
  Interface}\ }\textbf {\bibinfo {volume} {12}},\ \bibinfo {pages} {20150227}
  (\bibinfo {year} {2015})}\BibitemShut {NoStop}%
\bibitem [{\citenamefont {Ren}\ \emph {et~al.}(2017)\citenamefont {Ren},
  \citenamefont {Zhou}, \citenamefont {Mao}, \citenamefont {Xu}, \citenamefont
  {Huang},\ and\ \citenamefont {Mallouk}}]{ren2017rheotaxis}%
  \BibitemOpen
  \bibfield  {author} {\bibinfo {author} {\bibfnamefont {L.}~\bibnamefont
  {Ren}}, \bibinfo {author} {\bibfnamefont {D.}~\bibnamefont {Zhou}}, \bibinfo
  {author} {\bibfnamefont {Z.}~\bibnamefont {Mao}}, \bibinfo {author}
  {\bibfnamefont {P.}~\bibnamefont {Xu}}, \bibinfo {author} {\bibfnamefont
  {T.}~\bibnamefont {Huang}},\ and\ \bibinfo {author} {\bibfnamefont
  {T.}~\bibnamefont {Mallouk}},\ }\bibfield  {title} {\bibinfo {title}
  {Rheotaxis of bimetallic micromotors driven by chemical-acoustic hybrid
  power},\ }\href@noop {} {\bibfield  {journal} {\bibinfo  {journal} {ACS
  Nano}\ }\textbf {\bibinfo {volume} {11}},\ \bibinfo {pages} {10591 }
  (\bibinfo {year} {2017})}\BibitemShut {NoStop}%
\bibitem [{\citenamefont {Baker}\ \emph {et~al.}(2018)\citenamefont {Baker},
  \citenamefont {Kauffman}, \citenamefont {Laskar}, \citenamefont {Shklyaev},
  \citenamefont {Shum}, \citenamefont {Aronson}, \citenamefont {Balazs},\ and\
  \citenamefont {Sen}}]{rheotaxis2018baker}%
  \BibitemOpen
  \bibfield  {author} {\bibinfo {author} {\bibfnamefont {R.}~\bibnamefont
  {Baker}}, \bibinfo {author} {\bibfnamefont {J.}~\bibnamefont {Kauffman}},
  \bibinfo {author} {\bibfnamefont {A.}~\bibnamefont {Laskar}}, \bibinfo
  {author} {\bibfnamefont {O.}~\bibnamefont {Shklyaev}}, \bibinfo {author}
  {\bibfnamefont {H.}~\bibnamefont {Shum}}, \bibinfo {author} {\bibfnamefont
  {I.}~\bibnamefont {Aronson}}, \bibinfo {author} {\bibfnamefont
  {A.}~\bibnamefont {Balazs}},\ and\ \bibinfo {author} {\bibfnamefont
  {A.}~\bibnamefont {Sen}},\ }\bibfield  {title} {\bibinfo {title} {Fight the
  flow: collective rheotaxis of artificial swimmers in confined systems},\
  }\href@noop {} {\bibfield  {journal} {\bibinfo  {journal} {Nanoscale}\
  }\textbf {\bibinfo {volume} {11}},\ \bibinfo {pages} {10944} (\bibinfo {year}
  {2018})}\BibitemShut {NoStop}%
\bibitem [{\citenamefont {Wu}\ and\ \citenamefont
  {Libchaber}(2000)}]{WuLib2000}%
  \BibitemOpen
  \bibfield  {author} {\bibinfo {author} {\bibfnamefont {X.-L.}\ \bibnamefont
  {Wu}}\ and\ \bibinfo {author} {\bibfnamefont {A.}~\bibnamefont {Libchaber}},\
  }\bibfield  {title} {\bibinfo {title} {Particle {D}iffusion in a
  {Q}uasi-{T}wo-{D}imensional {B}acterial {B}ath},\ }\href@noop {} {\bibfield
  {journal} {\bibinfo  {journal} {Physical Review Letters}\ }\textbf {\bibinfo
  {volume} {84}},\ \bibinfo {pages} {3017} (\bibinfo {year}
  {2000})}\BibitemShut {NoStop}%
\bibitem [{\citenamefont {Vicsek}\ and\ \citenamefont
  {Zafeiris}(2012)}]{VisZaf2012}%
  \BibitemOpen
  \bibfield  {author} {\bibinfo {author} {\bibfnamefont {T.}~\bibnamefont
  {Vicsek}}\ and\ \bibinfo {author} {\bibfnamefont {A.}~\bibnamefont
  {Zafeiris}},\ }\bibfield  {title} {\bibinfo {title} {Collective motion},\
  }\href@noop {} {\bibfield  {journal} {\bibinfo  {journal} {Physics Reports}\
  }\textbf {\bibinfo {volume} {517}},\ \bibinfo {pages} {71} (\bibinfo {year}
  {2012})}\BibitemShut {NoStop}%
\bibitem [{\citenamefont {Ryan}\ \emph {et~al.}(5021)\citenamefont {Ryan},
  \citenamefont {Sokolov}, \citenamefont {Berlyand},\ and\ \citenamefont
  {Aranson}}]{RyaSokBerAra13}%
  \BibitemOpen
  \bibfield  {author} {\bibinfo {author} {\bibfnamefont {S.~D.}\ \bibnamefont
  {Ryan}}, \bibinfo {author} {\bibfnamefont {A.}~\bibnamefont {Sokolov}},
  \bibinfo {author} {\bibfnamefont {L.}~\bibnamefont {Berlyand}},\ and\
  \bibinfo {author} {\bibfnamefont {I.}~\bibnamefont {Aranson}},\ }\bibfield
  {title} {\bibinfo {title} {Correlation properties of collective motion in
  bacterial suspension},\ }\href@noop {} {\bibfield  {journal} {\bibinfo
  {journal} {New Journal of Physics, PMCID: PMC3878490}\ }\textbf {\bibinfo
  {volume} {15}},\ \bibinfo {pages} {105021} (\bibinfo {year} {2013.
  \url{http://dx.doi.org/10.1088/1367-2630/15/10/105021}})}\BibitemShut
  {NoStop}%
\bibitem [{\citenamefont {Vrugt}\ and\ \citenamefont
  {Wittkowski}(2024)}]{vrugt2024review}%
  \BibitemOpen
  \bibfield  {author} {\bibinfo {author} {\bibfnamefont {M.~t.}\ \bibnamefont
  {Vrugt}}\ and\ \bibinfo {author} {\bibfnamefont {R.}~\bibnamefont
  {Wittkowski}},\ }\bibfield  {title} {\bibinfo {title} {A review of active
  matter reviews},\ }\href@noop {} {\bibfield  {journal} {\bibinfo  {journal}
  {arXiv preprint arXiv:2405.15751}\ } (\bibinfo {year} {2024})}\BibitemShut
  {NoStop}%
\bibitem [{\citenamefont {Potomkin}\ \emph
  {et~al.}(2017{\natexlab{a}})\citenamefont {Potomkin}, \citenamefont {Kaiser},
  \citenamefont {Berlyand},\ and\ \citenamefont
  {Aranson}}]{potomkin2017focusing}%
  \BibitemOpen
  \bibfield  {author} {\bibinfo {author} {\bibfnamefont {M.}~\bibnamefont
  {Potomkin}}, \bibinfo {author} {\bibfnamefont {A.}~\bibnamefont {Kaiser}},
  \bibinfo {author} {\bibfnamefont {L.}~\bibnamefont {Berlyand}},\ and\
  \bibinfo {author} {\bibfnamefont {I.}~\bibnamefont {Aranson}},\ }\bibfield
  {title} {\bibinfo {title} {Focusing of active particles in a converging
  flow},\ }\href@noop {} {\bibfield  {journal} {\bibinfo  {journal} {New
  Journal of Physics}\ }\textbf {\bibinfo {volume} {19}},\ \bibinfo {pages}
  {115005} (\bibinfo {year} {2017}{\natexlab{a}})}\BibitemShut {NoStop}%
\bibitem [{\citenamefont {Rubio}\ \emph {et~al.}(2021)\citenamefont {Rubio},
  \citenamefont {Potomkin}, \citenamefont {Baker}, \citenamefont {Sen},
  \citenamefont {Berlyand},\ and\ \citenamefont {Aranson}}]{rubio2021self}%
  \BibitemOpen
  \bibfield  {author} {\bibinfo {author} {\bibfnamefont {L.~D.}\ \bibnamefont
  {Rubio}}, \bibinfo {author} {\bibfnamefont {M.}~\bibnamefont {Potomkin}},
  \bibinfo {author} {\bibfnamefont {R.~D.}\ \bibnamefont {Baker}}, \bibinfo
  {author} {\bibfnamefont {A.}~\bibnamefont {Sen}}, \bibinfo {author}
  {\bibfnamefont {L.}~\bibnamefont {Berlyand}},\ and\ \bibinfo {author}
  {\bibfnamefont {I.~S.}\ \bibnamefont {Aranson}},\ }\bibfield  {title}
  {\bibinfo {title} {Self-propulsion and shear flow align active particles in
  nozzles and channels},\ }\href@noop {} {\bibfield  {journal} {\bibinfo
  {journal} {Advanced Intelligent Systems}\ }\textbf {\bibinfo {volume} {3}},\
  \bibinfo {pages} {2000178} (\bibinfo {year} {2021})}\BibitemShut {NoStop}%
\bibitem [{\citenamefont {Drescher}\ \emph {et~al.}(2011)\citenamefont
  {Drescher}, \citenamefont {Dunkel}, \citenamefont {Cisneros}, \citenamefont
  {Ganguly},\ and\ \citenamefont {Goldstein}}]{drescher2011fluid}%
  \BibitemOpen
  \bibfield  {author} {\bibinfo {author} {\bibfnamefont {K.}~\bibnamefont
  {Drescher}}, \bibinfo {author} {\bibfnamefont {J.}~\bibnamefont {Dunkel}},
  \bibinfo {author} {\bibfnamefont {L.~H.}\ \bibnamefont {Cisneros}}, \bibinfo
  {author} {\bibfnamefont {S.}~\bibnamefont {Ganguly}},\ and\ \bibinfo {author}
  {\bibfnamefont {R.~E.}\ \bibnamefont {Goldstein}},\ }\bibfield  {title}
  {\bibinfo {title} {Fluid dynamics and noise in bacterial cell--cell and
  cell--surface scattering},\ }\href@noop {} {\bibfield  {journal} {\bibinfo
  {journal} {Proceedings of the National Academy of Sciences}\ }\textbf
  {\bibinfo {volume} {108}},\ \bibinfo {pages} {10940} (\bibinfo {year}
  {2011})}\BibitemShut {NoStop}%
\bibitem [{\citenamefont {Ezhilan}\ and\ \citenamefont
  {Saintillan}(2015)}]{EzhSai2015}%
  \BibitemOpen
  \bibfield  {author} {\bibinfo {author} {\bibfnamefont {B.}~\bibnamefont
  {Ezhilan}}\ and\ \bibinfo {author} {\bibfnamefont {D.}~\bibnamefont
  {Saintillan}},\ }\bibfield  {title} {\bibinfo {title} {Transport of a dilute
  active suspension in pressure-driven channel flow},\ }\href@noop {}
  {\bibfield  {journal} {\bibinfo  {journal} {Journal of Fluid Mechanics}\
  }\textbf {\bibinfo {volume} {777}},\ \bibinfo {pages} {480} (\bibinfo {year}
  {2015})}\BibitemShut {NoStop}%
\bibitem [{\citenamefont {Figueroa-Morales}\ \emph {et~al.}(2015)\citenamefont
  {Figueroa-Morales}, \citenamefont {Mino}, \citenamefont {Rivera},
  \citenamefont {Caballero}, \citenamefont {Cl\'ement}, \citenamefont
  {Altshuler},\ and\ \citenamefont {Lindner}}]{nuris2015}%
  \BibitemOpen
  \bibfield  {author} {\bibinfo {author} {\bibfnamefont {N.}~\bibnamefont
  {Figueroa-Morales}}, \bibinfo {author} {\bibfnamefont {G.}~\bibnamefont
  {Mino}}, \bibinfo {author} {\bibfnamefont {A.}~\bibnamefont {Rivera}},
  \bibinfo {author} {\bibfnamefont {R.}~\bibnamefont {Caballero}}, \bibinfo
  {author} {\bibfnamefont {E.}~\bibnamefont {Cl\'ement}}, \bibinfo {author}
  {\bibfnamefont {E.}~\bibnamefont {Altshuler}},\ and\ \bibinfo {author}
  {\bibfnamefont {A.}~\bibnamefont {Lindner}},\ }\bibfield  {title} {\bibinfo
  {title} {Living on the edge: transfer and traffic of {E}. {C}oli in a
  confined flow},\ }\href@noop {} {\bibfield  {journal} {\bibinfo  {journal}
  {Soft matter}\ }\textbf {\bibinfo {volume} {11}},\ \bibinfo {pages} {6284}
  (\bibinfo {year} {2015})}\BibitemShut {NoStop}%
\bibitem [{\citenamefont {Bukatin}\ \emph {et~al.}(2020)\citenamefont
  {Bukatin}, \citenamefont {Denissenko},\ and\ \citenamefont
  {Kantsler}}]{Bukatin2020}%
  \BibitemOpen
  \bibfield  {author} {\bibinfo {author} {\bibfnamefont {A.}~\bibnamefont
  {Bukatin}}, \bibinfo {author} {\bibfnamefont {P.}~\bibnamefont
  {Denissenko}},\ and\ \bibinfo {author} {\bibfnamefont {V.}~\bibnamefont
  {Kantsler}},\ }\bibfield  {title} {\bibinfo {title} {Self-organization and
  multi-line transport of human spermatozoa in rectangular microchannels due to
  cell-cell interactions},\ }\href@noop {} {\bibfield  {journal} {\bibinfo
  {journal} {Scientific reports}\ }\textbf {\bibinfo {volume} {10}},\ \bibinfo
  {pages} {9830} (\bibinfo {year} {2020})}\BibitemShut {NoStop}%
\bibitem [{\citenamefont {Nicolle}(2014)}]{nicolle2014}%
  \BibitemOpen
  \bibfield  {author} {\bibinfo {author} {\bibfnamefont {L.}~\bibnamefont
  {Nicolle}},\ }\bibfield  {title} {\bibinfo {title} {Catheter associated
  urinary tract infections},\ }\href@noop {} {\bibfield  {journal} {\bibinfo
  {journal} {Antimicrobial resistance and infection control}\ }\textbf
  {\bibinfo {volume} {3}},\ \bibinfo {pages} {1} (\bibinfo {year}
  {2014})}\BibitemShut {NoStop}%
\bibitem [{\citenamefont {Park}\ \emph {et~al.}(2003)\citenamefont {Park},
  \citenamefont {Wolanin}, \citenamefont {Yuzbashyan}, \citenamefont
  {Silberzan}, \citenamefont {Stock},\ and\ \citenamefont {Austin}}]{park2004}%
  \BibitemOpen
  \bibfield  {author} {\bibinfo {author} {\bibfnamefont {S.}~\bibnamefont
  {Park}}, \bibinfo {author} {\bibfnamefont {P.}~\bibnamefont {Wolanin}},
  \bibinfo {author} {\bibfnamefont {E.}~\bibnamefont {Yuzbashyan}}, \bibinfo
  {author} {\bibfnamefont {P.}~\bibnamefont {Silberzan}}, \bibinfo {author}
  {\bibfnamefont {J.}~\bibnamefont {Stock}},\ and\ \bibinfo {author}
  {\bibfnamefont {R.}~\bibnamefont {Austin}},\ }\bibfield  {title} {\bibinfo
  {title} {Motion to form a quorum},\ }\href@noop {} {\bibfield  {journal}
  {\bibinfo  {journal} {Science}\ }\textbf {\bibinfo {volume} {5630}},\
  \bibinfo {pages} {188} (\bibinfo {year} {2003})}\BibitemShut {NoStop}%
\bibitem [{\citenamefont {Figueroa-Morales}\ \emph {et~al.}(2020)\citenamefont
  {Figueroa-Morales}, \citenamefont {Rivera}, \citenamefont {Soto},
  \citenamefont {Lindner}, \citenamefont {Altshuler},\ and\ \citenamefont
  {Cl{\'e}ment}}]{figueroa2020coli}%
  \BibitemOpen
  \bibfield  {author} {\bibinfo {author} {\bibfnamefont {N.}~\bibnamefont
  {Figueroa-Morales}}, \bibinfo {author} {\bibfnamefont {A.}~\bibnamefont
  {Rivera}}, \bibinfo {author} {\bibfnamefont {R.}~\bibnamefont {Soto}},
  \bibinfo {author} {\bibfnamefont {A.}~\bibnamefont {Lindner}}, \bibinfo
  {author} {\bibfnamefont {E.}~\bibnamefont {Altshuler}},\ and\ \bibinfo
  {author} {\bibfnamefont {{\'E}.}~\bibnamefont {Cl{\'e}ment}},\ }\bibfield
  {title} {\bibinfo {title} {E. coli “super-contaminates” narrow ducts
  fostered by broad run-time distribution},\ }\href@noop {} {\bibfield
  {journal} {\bibinfo  {journal} {Science advances}\ }\textbf {\bibinfo
  {volume} {6}},\ \bibinfo {pages} {eaay0155} (\bibinfo {year}
  {2020})}\BibitemShut {NoStop}%
\bibitem [{\citenamefont {Ramia}\ \emph {et~al.}(1993)\citenamefont {Ramia},
  \citenamefont {Tullock},\ and\ \citenamefont {Phan-Thien}}]{RamTulPha1993}%
  \BibitemOpen
  \bibfield  {author} {\bibinfo {author} {\bibfnamefont {M.}~\bibnamefont
  {Ramia}}, \bibinfo {author} {\bibfnamefont {D.}~\bibnamefont {Tullock}},\
  and\ \bibinfo {author} {\bibfnamefont {N.}~\bibnamefont {Phan-Thien}},\
  }\bibfield  {title} {\bibinfo {title} {The role of hydrodynamic interaction
  in the locomotion of microorganisms},\ }\href@noop {} {\bibfield  {journal}
  {\bibinfo  {journal} {Biophysical Journal}\ }\textbf {\bibinfo {volume}
  {65}},\ \bibinfo {pages} {755} (\bibinfo {year} {1993})}\BibitemShut
  {NoStop}%
\bibitem [{\citenamefont {DiLuzio}\ \emph {et~al.}(2005)\citenamefont
  {DiLuzio}, \citenamefont {Turner}, \citenamefont {Mayer}, \citenamefont
  {Garstecki}, \citenamefont {Weibel}, \citenamefont {Berg},\ and\
  \citenamefont {Whitesides}}]{diluzio2005}%
  \BibitemOpen
  \bibfield  {author} {\bibinfo {author} {\bibfnamefont {W.}~\bibnamefont
  {DiLuzio}}, \bibinfo {author} {\bibfnamefont {L.}~\bibnamefont {Turner}},
  \bibinfo {author} {\bibfnamefont {M.}~\bibnamefont {Mayer}}, \bibinfo
  {author} {\bibfnamefont {P.}~\bibnamefont {Garstecki}}, \bibinfo {author}
  {\bibfnamefont {D.}~\bibnamefont {Weibel}}, \bibinfo {author} {\bibfnamefont
  {H.}~\bibnamefont {Berg}},\ and\ \bibinfo {author} {\bibfnamefont
  {G.}~\bibnamefont {Whitesides}},\ }\bibfield  {title} {\bibinfo {title} {{\it
  Escherichia coli} swim on the right-hand side},\ }\href@noop {} {\bibfield
  {journal} {\bibinfo  {journal} {Nature}\ }\textbf {\bibinfo {volume} {435}},\
  \bibinfo {pages} {1271} (\bibinfo {year} {2005})}\BibitemShut {NoStop}%
\bibitem [{\citenamefont {Potomkin}\ \emph
  {et~al.}(2017{\natexlab{b}})\citenamefont {Potomkin}, \citenamefont
  {Tournus}, \citenamefont {Berlyand},\ and\ \citenamefont
  {Aranson}}]{potomkin2017flagella}%
  \BibitemOpen
  \bibfield  {author} {\bibinfo {author} {\bibfnamefont {M.}~\bibnamefont
  {Potomkin}}, \bibinfo {author} {\bibfnamefont {M.}~\bibnamefont {Tournus}},
  \bibinfo {author} {\bibfnamefont {L.}~\bibnamefont {Berlyand}},\ and\
  \bibinfo {author} {\bibfnamefont {I.}~\bibnamefont {Aranson}},\ }\bibfield
  {title} {\bibinfo {title} {Flagella bending affects macroscopic properties of
  bacterial suspensions},\ }\href@noop {} {\bibfield  {journal} {\bibinfo
  {journal} {Journal of The Royal Society Interface}\ }\textbf {\bibinfo
  {volume} {14}},\ \bibinfo {pages} {20161031} (\bibinfo {year}
  {2017}{\natexlab{b}})}\BibitemShut {NoStop}%
\bibitem [{\citenamefont {Berlyand}\ \emph {et~al.}(2020)\citenamefont
  {Berlyand}, \citenamefont {Jabin}, \citenamefont {Potomkin},\ and\
  \citenamefont {Ratajczyk}}]{potomkin2020confined}%
  \BibitemOpen
  \bibfield  {author} {\bibinfo {author} {\bibfnamefont {L.}~\bibnamefont
  {Berlyand}}, \bibinfo {author} {\bibfnamefont {P.}~\bibnamefont {Jabin}},
  \bibinfo {author} {\bibfnamefont {M.}~\bibnamefont {Potomkin}},\ and\
  \bibinfo {author} {\bibfnamefont {E.}~\bibnamefont {Ratajczyk}},\ }\bibfield
  {title} {\bibinfo {title} {A kinetic approach to active rods dynamics in
  confined domains},\ }\href@noop {} {\bibfield  {journal} {\bibinfo  {journal}
  {Multiscale Modeling \& Simulation}\ }\textbf {\bibinfo {volume} {18}},\
  \bibinfo {pages} {1} (\bibinfo {year} {2020})}\BibitemShut {NoStop}%
\end{thebibliography}%

%%%%%%%%%%%%%%%%%%%%%%%%%%%%%%%%%%%%%%%%%%
% SUPPLEMENTARY INFORMATION
%%%%%%%%%%%%%%%%%%%%%%%%%%%%%%%%%%%%%%%%%%
\newpage 

\onecolumngrid

\setcounter{equation}{0}

\renewcommand{\theequation}{SI-\arabic{equation}}
\centerline{\bf \large Supplementary Information}

%\tableofcontents

  \section{Supplementary Videos}

\noindent Supplementary videos ({\it SI Videos}) are available online: \url{https://sites.google.com/view/mpotomkin/simulation-videos}.

\smallskip

\noindent{\bf Supplementary Video 1.} Monte Carlo simulations for individual active rods swimming in the cylindrical microchannel with the elliptic cross-section. The ellipse has semi-major axis $a=4$ and semi-minor axis $b=1$. The left half of video frames is the 3D depiction of the active rod dynamics. An active rod is a red segment with the green front. The wall is light blue. The downward black arrow shows the direction of background flow. All rods start from $z=0$. The upper right part of the video shows dynamics of active rods' centers (red points) projected on the $xy$-plane. The lower right part presents plots showing the fraction of rods accumulated at the walls, and at regions of high and low curvature. A frame from this video is used for Figure~2 of the main text. The video demonstrates that active rods tend to swim upstream and concentrate at locations of high curvature.

\smallskip

\noindent{\bf Supplementary Video 2.} Monte Carlo simulations for individual active rods and three different flow rates $|w(0,0)|$: 0.1, 0.5, and 1.5. Active rods are drawn as red segments with green fronts. The video illustrates how the spreading of active rod density profiles depends on the background flow rates. Namely, the upstream front is flat and moves with the speed independent of the background flow rate. The downstream front is parabolic and its speed grows with the background flow rate.       

\smallskip 

\noindent{\bf Supplementary Video 3.} Monte Carlo simulations of individual active rods accounting for wall torque are presented. In the upper half of the video, the 3D dynamics of the active rods are shown. The microchannel is oriented horizontally, with active rods represented as red segments with a green front. In the lower half of the video, projections of the active rods onto the $xy$-plane are displayed. When an active rod swims along the microchannel, it appears as a green point in the lower part of the video. Conversely, when an active rod swims across the microchannel, it appears as a long red segment. The video illustrates that active rods concentrate at the locations of highest curvature and then reorient to swim along the microchannel. It also demonstrates that active rods are transported along the wall more quickly and detach from the wall more often compared to simulations that do not account for wall torque.

\smallskip 

\noindent{\bf Supplementary Video 4.} Monte Carlo simulations of individual active rods in the 2D lake problem, without wall torque. There is no background flow or rotational diffusion. In this scenario, an active rod does not reorient, so it collides with the wall and swims along the wall until it reaches a point where the rod is exactly perpendicular to the wall. An active rod is depicted as a red segment with the green front and black center. The setup for this video is the same as that used for deriving the active rod distribution $f(\alpha)$ in formula (14) of the main text.     

\smallskip 

\noindent{\bf Supplementary Video 5.} Monte Carlo simulations of individual active rods in the 2D lake problem, with wall torque. The wall torque coefficient $\mathcal{T}=0.1$ is small so there is a trapping region at high curvature points where all active rods eventually accumulate. This video illustrates the bifurcation phenomenon, that is, there is a threshold curvature so that orientation $\boldsymbol{p}=\boldsymbol{n}$ is unstable for sub-critical curvatures and is stable for super-critical curvatures. Recall that $\boldsymbol{p}$ and $\boldsymbol{n}$ are the orientation of the active rod and the outward normal vector at the wall location where the active rod touches the wall. If $\boldsymbol{p}=\boldsymbol{n}$, the active rod does not move. Instability of $\boldsymbol{p}=\boldsymbol{n}$ implies that the active rod will swim away from the current wall location. Note the difference between this video and SI Video 4: in both videos, active rods accumulate at high curvature locations. However, in SI Video 4, active rods can, though with a small probability, end up at a low curvature location, whereas in the current video, no active rods are found away from high curvature locations.     

\smallskip

\noindent{\bf Supplementary Video 6.} Monte Carlo simulations of individual active rods in the 2D circular lake, with sub-critical and super-critical wall torque coefficients $\mathcal{T}$. An active rod is depicted as a red segment with the green front and black center. In the sub-critical case (left), active rods eventually slow down and stop in a position perpendicular to the wall curve. In the super-critical case (right), there is a limiting angle $\varphi_\infty$ such that the angle between an active rod and the wall converges to $\pm \varphi_\infty$, and all active rods swim along the wall, either clockwise or counter-clockwise, at the same speed.     

\smallskip

\noindent{\bf Supplementary Video 7.}  Monte Carlo simulations for individual active rods in a rectangular microchannel. Active rods are initiated at $z=0$ (note the axis $z$ is oriented horizontally in the video) with random orientation and $(x,y)$-location. The video illustrates the four-lane structure of the active rod transport, that is, the active rods accumulate at corners.

	\section{Model of the active rod at the wall: the wall torque}
	Our main objective in this supplementary section is to derive the term in (13) of the main text of the manuscript. This term corresponds to the wall torque. We will repeat some steps from the manuscript but we will add the wall force. Also, we will consider below $\mathcal{T}=1$.
 
 First, write force and torque balances for the rod: 
	\begin{eqnarray}
	&&	-\eta \int\limits_{-\ell/2}^{\ell/2} \boldsymbol{V}(s) \,\text{d}s +V_{\text{prop}}\boldsymbol{p} + \boldsymbol{F}_{\text{wall}}=\boldsymbol{0},\label{eqn:force}\\
	&& -\eta\int\limits_{-\ell/2}^{\ell/2} s\boldsymbol{p}\times \boldsymbol{V}(s)\,\text{d}s + \dfrac{\ell}{2}\boldsymbol{p}\times \boldsymbol{F}_{\text{wall}}=\boldsymbol{0}.\label{eqn:torque}
	\end{eqnarray}
	We define the relative velocity $\boldsymbol{V}(s)$ as follows
	\begin{equation}
		\boldsymbol{V}(s)=\dfrac{\text{d}}{\text{d}t}\left[\boldsymbol{r}(t)+s\boldsymbol{p}(t)\right]-\boldsymbol{u}(\boldsymbol{r}(t)+s\boldsymbol{p}(t)). 
	\end{equation}
	Due to that $-\ell/2<s<\ell/2$ and smallness of rod's length $\ell$, we can approxiamte $\boldsymbol{V}(s)$:
	\begin{equation}
		\boldsymbol{V}(s)\approx \dot{\boldsymbol{r}}+s\dot{\boldsymbol{p}} - \boldsymbol{u}(\boldsymbol{r})-s(\nabla \boldsymbol{u}(\boldsymbol{r}))\boldsymbol{p}.\label{approx-for-V}
	\end{equation}
	By substituting \eqref{approx-for-V} into the force balance we get 
	\begin{equation}
		 \dot{\boldsymbol{r}}= \boldsymbol{u}(\boldsymbol{r})+\dfrac{F_{\text{prop}}}{\eta \ell}\boldsymbol{p}+\dfrac{1}{\eta\ell}\boldsymbol{F}_{\text{wall}}.\label{eqn_for_r}
	\end{equation} 
	At this point, we can find an expression for $\boldsymbol{F}_{\text{wall}}$ from the non-penetration condition $\dot{\boldsymbol{r}}\cdot \boldsymbol{n}=0$ provided that $\boldsymbol{n}$ is the outward normal vector on the wall and the rod is at the wall such that $\dot{\boldsymbol{r}}\cdot \boldsymbol{n}>0$:
	\begin{equation}
		\boldsymbol{F}_{\text{wall}} = - \boldsymbol{n}\boldsymbol{n}^{\text{T}}(\eta\ell\boldsymbol{u}(\boldsymbol{r})+F_{\text{prop}}\boldsymbol{p}).
		\end{equation}
		(Here, we simplify our calculation by assuming smallness of $\ell$. The exact condition of non-penetration for the front is $(\dot{\boldsymbol{r}}+\dfrac{\ell}{2}\dot{\boldsymbol{p}})\cdot \boldsymbol{n}=0$.  )
		
		The backgound flow vanishes at the wall due to no-slip boundary conditions, $\boldsymbol{u}|_{\text{wall}}=\boldsymbol{0}$. Thus the expression for the wall force $\boldsymbol{F}_{\text{wall}}$ reduces to 
		\begin{equation}
			\boldsymbol{F}_{\text{wall}} = -F_{\text{prop}}\boldsymbol{n}\boldsymbol{n}^{\text{T}}\boldsymbol{p}.\label{eqn:expression-for-F_prop}
			\end{equation}

	Next, we write the torque balance using \eqref{approx-for-V} and \eqref{eqn:expression-for-F_prop}:
	\begin{equation}
		\dfrac{\eta\ell^3}{12}\boldsymbol{p}\times\dot{\boldsymbol{p}} = \dfrac{\eta\ell^3}{12}\boldsymbol{p}\times ((\nabla \boldsymbol{u}(\boldsymbol{r}))\boldsymbol{p}) - \dfrac{\ell F_{\text{prop}}}{2}\boldsymbol{p}\times (\boldsymbol{n}(\boldsymbol{n}^{\text{T}}\boldsymbol{p}))
	\end{equation}
	Rewrite this expression using the identity $\boldsymbol{p}\times (\boldsymbol{p}\times \boldsymbol{v})=-(\text{I} - \boldsymbol{p}\boldsymbol{p}^{\text{T}})\boldsymbol{v}$ implying that $\boldsymbol{p}\times (\boldsymbol{p}\times \boldsymbol{v})$ is the minus orthogonal complement to the projection of the vector $\boldsymbol{v}$ onto $\boldsymbol{p}$.  For example, if  we substitute $\boldsymbol{v}=\dot{\boldsymbol{p}}$, then $\boldsymbol{p}\times (\boldsymbol{p}\times \dot{\boldsymbol{p}})=-\dot{\boldsymbol{p}}$ since $\dot{\boldsymbol{p}}$ is orthogonal to $\boldsymbol{p}$. Thus, we obtain the equation for $\boldsymbol{p}$:
	\begin{equation}
		\dot{\boldsymbol{p}}=\left(\text{I}-\boldsymbol{p}\boldsymbol{p}^{\text{T}}\right)((\nabla \boldsymbol{u}(\boldsymbol{r}))\boldsymbol{p})-\dfrac{6 F_{\text{prop}}}{\eta\ell^2}(\boldsymbol{n}^{\text{T}}\boldsymbol{p})\left(\text{I}-\boldsymbol{p}\boldsymbol{p}^{\text{T}}\right)\boldsymbol{n}.\label{eqn_for_p}
		\end{equation}
	Using $V_{\text{prop}}=F_{\text{prop}}/(\eta\ell)$ we get that the second in the right-hand side of \eqref{eqn_for_p} is 
 \begin{equation}
 -\dfrac{6 V_{\text{prop}}}{\ell}(\boldsymbol{n}^{\text{T}}\boldsymbol{p})\left(\text{I}-\boldsymbol{p}\boldsymbol{p}^{\text{T}}\right)\boldsymbol{n}.
 \end{equation}
 
	\begin{remark}
		We assume that the drag force is proportional to velocity $\boldsymbol{V}(s)$, as in the Stokes law. However, it is more reasonable to use the slender body theory and assume that the drag coefficients along and perpendicular to the rod are different. Namely, the drag force then is 
		\begin{equation}
			-\eta_\perp\int\limits_{-\ell/2}^{\ell/2} \boldsymbol{p}\boldsymbol{p}^{\mathrm{T}}\boldsymbol{V}(s)\,\mathrm{d}s -\eta_\parallel\int\limits_{-\ell/2}^{\ell/2} (\mathrm{I}-\boldsymbol{p}\boldsymbol{p}^{\mathrm{T}})\boldsymbol{V}(s)\,\mathrm{d}s  \label{slender-body-theory-expression}
			\end{equation}
			Substituting \eqref{approx-for-V}  and introducing the matrix 
			\begin{equation}
			M_\eta = \eta_\perp \ell \boldsymbol{p}\boldsymbol{p}^{\mathrm{T}}+ \eta_{\parallel}\ell(\mathrm{I}-\boldsymbol{p}\boldsymbol{p}^{\mathrm{T}})=\eta_\parallel\ell(\mathrm{I}-(1-\dfrac{\eta_{\perp}}{\eta_{\parallel}})\boldsymbol{p}\boldsymbol{p}^{\mathrm{T}})
			\end{equation}
			we rewrite \eqref{slender-body-theory-expression} and \eqref{eqn_for_r} as 
			\begin{equation}
			 \dot{\boldsymbol{r}}= \boldsymbol{u}(\boldsymbol{r})+F_{\mathrm{prop}}M^{-1}_{\eta}\boldsymbol{p}+M^{-1}_{\eta}\boldsymbol{F}_{\mathrm{wall}}.
			\end{equation} 
			Note that $M_{\eta}\boldsymbol{p} = \eta_{\perp}\boldsymbol{p}$ and thus
%			p = x - a * p (p^T x)
\begin{equation}
	\dot{\boldsymbol{r}}= \boldsymbol{u}(\boldsymbol{r})+\dfrac{F_{\mathrm{prop}}}{\eta_{\perp}\ell}\boldsymbol{p}+M^{-1}_{\eta}\boldsymbol{F}_{\mathrm{wall}}.
\end{equation} 
So the definition of $\boldsymbol{F}_{\text{wall}}$ is
\begin{equation}
\boldsymbol{F}_{\text{wall}} = - \dfrac{F_{\text{prop}}}{\eta_\perp\ell}(\boldsymbol{n}^{\text{T}}\boldsymbol{p})M_{\eta}\boldsymbol{n}.
\end{equation}
Now, let us examine what happens with the visous torque: 
\begin{equation}
	-\eta_\perp\int\limits_{-\ell/2}^{\ell/2} s\boldsymbol{p}\times\boldsymbol{p}\boldsymbol{p}^{\text{T}}\boldsymbol{V}(s)\,\mathrm{d}s -\eta_\parallel\int\limits_{-\ell/2}^{\ell/2} s\boldsymbol{p}\times(\text{I}-\boldsymbol{p}\boldsymbol{p}^{\text{T}})\boldsymbol{V}(s)\,\mathrm{d}s  \label{slender-body-theory-expression-torque}
	\end{equation}
Terms of the form $\boldsymbol{p}\times\boldsymbol{p}\boldsymbol{p}^{\text{T}}\boldsymbol{V}(s)=(\boldsymbol{p}^{\text{T}}\boldsymbol{V}(s))\boldsymbol{p}\times\boldsymbol{p}$ vanish and we have that the viscous torque \eqref{slender-body-theory-expression-torque} is simply
\begin{equation}
	-\eta_{\parallel}\int\limits_{-\ell/2}^{\ell/2}s\boldsymbol{p}\times \boldsymbol{V}(s)\,\mathrm{d}s.
	\end{equation}
	So the torque equation becomes 
		\begin{equation}
		\dot{\boldsymbol{p}}=\left(\mathrm{I}-\boldsymbol{p}\boldsymbol{p}^{\text{T}}\right)((\nabla \boldsymbol{u}(\boldsymbol{r}))\boldsymbol{p})-\dfrac{6 F_{\text{prop}}}{\eta_{\perp}\eta_{\parallel}\ell^3}(\boldsymbol{n}^{\text{T}}\boldsymbol{p})\left(\mathrm{I}-\boldsymbol{p}\boldsymbol{p}^{\text{T}}\right)M_{\eta}\boldsymbol{n}.
	\end{equation}
	We note that $\left(\mathrm{I}-\boldsymbol{p}\boldsymbol{p}^{\text{T}}\right)M_{\eta}=\eta_{\parallel}\ell (\mathrm{I}-\boldsymbol{p}\boldsymbol{p}^{\mathrm{T}})$ so we conclude that the torque balance here is reduced to \eqref{eqn_for_p} with $\eta=\eta_{\perp}$. 
		\end{remark}
	
	\section{Elliptic lake problem}
	The main objective of this suppementary section is to derive the formula (14) and the equation (15) in the main text of the manuscript. 
 
 Consider two-dimensional elliptic domain $\Omega=\left\{x^2/a^2+y^2/b^2<1\right\}$ with no background flow $\boldsymbol{u}=\boldsymbol{0}$ so there is no current and thus we will call $\Omega$ a lake. Parameterize the boundary (wall) of the lake with angle $\alpha$ such that 
	\begin{equation}
		\partial \Omega: \left\{
		\begin{array}{l}
		x=a\cos(\alpha),\\
		y=b\sin(\alpha),
		\end{array}
		\quad 0\leq \alpha < 2\pi.
		\right.
	\end{equation} 

 It will be convenient to have an expression for the tangent and normal vectors at the wall: 
	\begin{equation}
		\boldsymbol{\tau} = \dfrac{1}{J(\alpha)}
		\left[
		\begin{array}{c}
		-a\sin(\alpha)\\b\cos(\alpha)
		\end{array}
		\right],\,
		\boldsymbol{n} = \dfrac{1}{J(\alpha)}
		\left[
		\begin{array}{c}
			b\cos(\alpha)\\a\sin(\alpha)
		\end{array}
		\right],\,
		J(\alpha)=\sqrt{b^2\cos^2(\alpha)+a^2\sin^2(\alpha)}.
		\end{equation}
  
\subsection{Large-time wall distribution of active rods.} Assume that rods do not experience any wall torque and the initial orientation of rods is uniform $\varphi\sim \mathcal{U}(0,2\pi)$. In this case, the rod's orientation will not change as time evolves since we disregard all forces that may reorient the rod: the background shear, rotational diffusion, and the wall torque. The rod will swim straight until it collides with the wall, and then it will slide along the wall until its orientation vector $\boldsymbol{p}$ coincides with the outward normal vector $\boldsymbol{n}$: 
\begin{equation}
 \dfrac{1}{J(\alpha)}
		\left[
		\begin{array}{c}
			b\cos(\alpha)\\a\sin(\alpha)
		\end{array}
		\right] = \left[\begin{array}{l}
    \cos(\varphi)\\
    \sin(\varphi)
  \end{array}\right].\label{p_equals_n}
\end{equation}
Our task here is to find the probability distribution function for $\alpha$ which we denote by $f_{\alpha}(\alpha)$ from \eqref{p_equals_n}. We find this function by considering first $0\leq \alpha<\pi$ so we have $\varphi = \varphi(\alpha)= \arccos({b\cos(\alpha})/J(\alpha))$. Using the relation between cumulative and probability distribution functions we get
\begin{eqnarray}
f_{\alpha}(\hat{\alpha}) &=& \dfrac{\partial}{\partial_{\hat{\alpha}}}\left[\mathbb P(0\leq \alpha\leq \hat{\alpha})\right]\nonumber\\
&=& \dfrac{1}{2\pi}\dfrac{\partial}{\partial_{\hat{\alpha}}}\left[\mathbb P(0\leq \varphi\leq \varphi(\hat{\alpha}))\right]\nonumber\\
&=& \dfrac{1}{2\pi}\dfrac{\partial}{\partial_{\hat{\alpha}}}\left[\arccos\left(\frac{b}{J(\hat{\alpha})}\cos(\hat{\alpha})\right)\right]\nonumber \\ 
&=& -\dfrac{1}{2\pi}\dfrac{1}{\sqrt{1- \dfrac{b^2\cos^2(\hat\alpha)}{J^2(\hat\alpha)}}}\left(\dfrac{-b\sin(\hat\alpha)}{J(\hat\alpha)}-\dfrac{b(a^2-b^2)\cos^2(\hat\alpha)\sin(\hat\alpha)}{J^3(\hat\alpha)}\right)
\nonumber\\
&=& \dfrac{1}{2\pi}\dfrac{J(\hat\alpha)}{a\sin(\hat\alpha)}\cdot b\sin(\hat\alpha)\cdot \dfrac{J^2(\hat\alpha)+(a^2-b^2)\cos^2(\hat\alpha)}{J^3(\hat\alpha)}
\nonumber \\ 
&=& \dfrac{ab}{2\pi(b^2\cos^2(\hat{\alpha})+a^2\sin^2(\hat{\alpha}))}.
\end{eqnarray}
Now, dropping hats in notations for $\alpha$ and extending the definition for $f_{\alpha}$ by symmetry $f_{\alpha}(\alpha) = f_{\alpha}(2 \pi - \alpha)$ we can finally conclude that 
\begin{equation}
f_{\alpha}(\alpha) = \dfrac{ab}{2\pi(b^2\cos^2(\alpha)+a^2\sin^2(\alpha))}.
\end{equation}
This function attains minimal value $\dfrac{b}{2\pi a}$ at $\alpha=\frac{\pi}{2}$ and $\alpha=\frac{3\pi}{2}$ and maximal value $\dfrac{a}{2\pi b}$ at $\alpha = 0$ and $\alpha = \pi$.

 \subsection{Large-time wall distribution in case of the wall torque}
 
 Now, we will assume that a rod is attached to the wall and its front location is described by the function $\alpha(t)$ meaning that the location of the front is $M:=(a \cos(\alpha(t)),b\sin(\alpha(t)))$. The direction of the rod is given by orientation angle $\varphi(t)$ so $\boldsymbol{p}=(\cos(\varphi(t)),\sin(\varphi(t)))^{\text{T}}$. Below, we derive equations for $\alpha(t)$ and $\varphi(t)$.

	Next, rewrite \eqref{eqn_for_r} and \eqref{eqn_for_p} as 
	\begin{equation}
		\left\{
		\begin{array}{l}
			\dot{\boldsymbol{r}} = V_{\text{prop}}(\mathrm{I}-\boldsymbol{n}\boldsymbol{n}^{\mathrm{T}})\boldsymbol{p},
			\\
			\dot{\boldsymbol{p}} = -\mathcal{T}\dfrac{6 V_{\text{prop}}}{\ell}(\boldsymbol{n}^{\text{T}}\boldsymbol{p})\left(\text{I}-\boldsymbol{p}\boldsymbol{p}^{\text{T}}\right)\boldsymbol{n}.
		\end{array}
		\right.\label{eqn:lake-system}
		\end{equation}
		Here $V_{\text{prop}}=(\eta \ell)^{-1}F_{\text{prop}}$. These equations can be rewritten in a more concise form noting that $(\mathrm{I}-\boldsymbol{n}\boldsymbol{n}^{\mathrm{T}})\boldsymbol{p}=(\boldsymbol{p}\cdot\boldsymbol{\tau})\boldsymbol{\tau}$ and $\left(\text{I}-\boldsymbol{p}\boldsymbol{p}^{\text{T}}\right)\boldsymbol{n}= (\boldsymbol{n}\cdot\boldsymbol{e}_\varphi)\boldsymbol{e}_\varphi$ where $\boldsymbol{e}_\varphi=\partial_\varphi \boldsymbol{p}$ is a unit vector orthogonal to $\boldsymbol{p}$:
        \begin{equation}
		\left\{
		\begin{array}{l}
			\dot{\boldsymbol{r}} = V_{\text{prop}}(\boldsymbol{p}\cdot\boldsymbol{\tau})\boldsymbol{\tau},
			\\
			\dot{\boldsymbol{p}} = -\mathcal{T}\dfrac{6 V_{\text{prop}}}{\ell}(\boldsymbol{n}\cdot\boldsymbol{p})(\boldsymbol{n}\cdot\boldsymbol{e}_\varphi)\boldsymbol{e}_\varphi.
		\end{array}
		\right.\label{eqn:lake-system-concise}
		\end{equation}
		
	\noindent{\it Equation for $\alpha$}. 
	The equation for $\alpha$ is obtained from that the rod is on the wall $\boldsymbol{r}=\left[a \cos(\alpha),b\sin(\alpha)\right]^{\text{T}}$ and the equation for $\boldsymbol{r}$ in \eqref{eqn:lake-system-concise}:  
	\begin{equation}
		\dot{\alpha} = V_{\mathrm{prop}}(J(\alpha))^{-1} (\boldsymbol{p}\cdot\boldsymbol{\tau}).
		\end{equation}
		\noindent{\it Equation for $\varphi$.} First, observe that $\dot{\boldsymbol{p}}=\dot{\varphi}\boldsymbol{e}_\varphi$, where
		\begin{equation}
				\boldsymbol{e}_{\varphi} =
			\left[
			\begin{array}{c}
				-\sin(\varphi)\\\cos(\varphi)
			\end{array}
			\right].
		\end{equation}
		Then the equation for $\boldsymbol{p}$ in \eqref{eqn:lake-system-concise} becomes
		\begin{equation}
			\dot{\varphi}=-\mathcal{T}\dfrac{6V_{\mathrm{prop}}}{\ell}(\boldsymbol{p}\cdot \boldsymbol{n})(\boldsymbol{n}\cdot\boldsymbol{e}_{\varphi}). 
		\end{equation} 
	Therefore, we obtain the system for angles $\alpha$ and $\varphi$. Introduce also the angle between vectors $\boldsymbol{n}$ and $\boldsymbol{p}$. 
 
     \begin{wrapfigure}[12]{r}{0.33\textwidth}
  \begin{center}
    \includegraphics[width=0.33\textwidth]{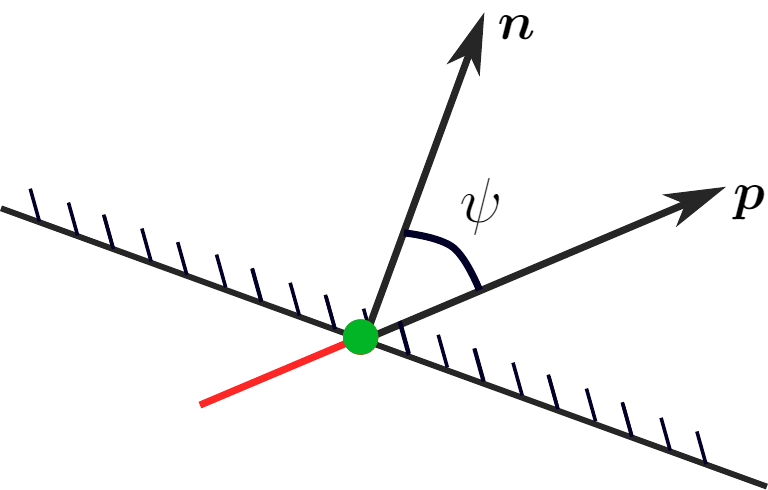}
  \end{center}
\end{wrapfigure}
 Specifically, the angle $\psi$ is defined by
	\begin{equation}
		\cos(\psi)=\boldsymbol{p}\cdot \boldsymbol{n} \text{ and }
		\sin(\psi)=\boldsymbol{p}\cdot \boldsymbol{\tau}.
		\end{equation}
		Differentiate in time the identity $\cos(\psi)=\boldsymbol{p}\cdot \boldsymbol{n}$. Note that one can get directly from the chain rule that $\partial_t \cos(\psi)=-\dot{\psi}(\boldsymbol{p}\cdot \boldsymbol{\tau})$. Also, direct differentiation of $\boldsymbol{n}$ gives
		\begin{equation}
			\dot{\boldsymbol{n}} = -\dfrac{\partial_{\alpha}J}{J}\boldsymbol{n}+\left[\begin{array}{c}-b\sin(\alpha)\\a\cos(\alpha)\end{array}\right]\dfrac{\dot{\alpha}}{J}
		\end{equation}
		From $\dot{\boldsymbol{n}}\cdot\boldsymbol{n}=0$ we get $\dot{\boldsymbol{n}}=(\dot{\boldsymbol{n}}\cdot \boldsymbol{\tau})\boldsymbol{\tau}$ and hence 
		\begin{equation}
		\dot{\boldsymbol{n}}=\dfrac{ab}{J^2}\dot{\alpha}\boldsymbol{\tau}.
			\end{equation}
		Now differentiation of the right-hand side of $\cos(\psi)=\boldsymbol{p}\cdot \boldsymbol{n}$ gives 
		\begin{eqnarray}
			-\dot{\psi}(\boldsymbol{p}\cdot \boldsymbol{\tau})&=& \dot{\boldsymbol{p}}\cdot \boldsymbol{n} + \boldsymbol{p}\cdot \dot{\boldsymbol{n}}\nonumber\\
			&=& \dot{\varphi}(\boldsymbol{e}_\varphi\cdot \boldsymbol{n}) +\dfrac{ab}{J^2}\dot{\alpha}(\boldsymbol{p}\cdot \boldsymbol{\tau})\label{eqn:relation-between-angles}
	   \end{eqnarray} 
		Finally, observe that $(\boldsymbol{p}\cdot \boldsymbol{\tau})=-(\boldsymbol{e}_\varphi\cdot \boldsymbol{n})$ so we have 
		\begin{equation}
			\dot{\psi} = \dot{\varphi} - \dfrac{ab}{J^2}\dot{\alpha}.
			\end{equation}
		This equation allows us to write the system for $\alpha$ and $\psi$: 
		\begin{equation}
			\left\{
			\begin{array}{l}
				\dot{\alpha}=V_{\mathrm{prop}}J^{-1} \sin(\psi)\\
				\dot{\psi} = \dfrac{6V_{\mathrm{prop}}}{\ell} \sin\psi\left(\mathcal{T}\cos(\psi) - \dfrac{ab\ell}{6J^3}\right).
				\end{array}
			\right.
			\end{equation}
		Consider a particular case of this system that allows for decoupling. Namely, take $a=b=R$, then $J\equiv R$, and we can consider equation for $\psi$ only
  \begin{equation} 
  \dot{\psi} = \dfrac{6V_{\mathrm{prop}}}{\ell} \sin\psi\left(\mathcal{T}\cos(\psi) - \dfrac{\ell}{6R}\right).
  \end{equation}

\end{document}